\documentclass[a4paper,11pt]{article}
\pdfoutput=1 

\usepackage{jheppub}

\usepackage{lineno}
\usepackage[utf8]{inputenc}
\usepackage{amsmath}
\usepackage{amssymb}
\usepackage{amsfonts}
\usepackage{amsthm}
\usepackage{xcolor}
\usepackage{hyperref}
\usepackage{rotating}
\usepackage{array}
\usepackage{appendix}
\usepackage{bbold}
\usepackage{textcomp}
\usepackage{float}

\usepackage{tikz}
\usetikzlibrary{arrows,chains,matrix,positioning,scopes}

\def\kbar{{\mathchar'26\mkern-9muk}}

\def\de{\mathrm{d}}
\def\diff{\mathrm{d}}
\def\dj{d\kern-0.4em\char"16\kern-0.1em}
\def \Dj {\mbox{\raise0.3ex\hbox{-}\kern-0.4em D}}
\def\t{{\tt t}}
\def\tx{{\tt x}}

\def\tu{{\tt u}}
\def\tz{{\tt z}}
\def\te{{\tt e}}
\def\tg{{\tt g}}

\def\hx{{\hat x}}

\def\hz{{\hat z}}
\def\htt{{\hat t}}
\def\hp{{\hat p}}

\def\DDelta{{{\mathbb{\Delta}}}}

\def\p{{\partial}}

\title{QFT on Fuzzy AdS Spaces:\\
Classical Limit and Boundary Correlation Functions}

\author[a]{Bojana Brki\'c,}
\author[b]{Ilija Buri\'c,}
\author[a]{Maja Buri\'c,}
\author[a]{Du\v san \Dj or\dj evi\'c}
\author[a]{and Du\v sko Latas}

\affiliation[a]{Faculty of Physics, University of Belgrade, Studentski trg 12, 11001 Belgrade, Serbia}
\affiliation[b]{School of Mathematics and Hamilton Mathematics Institute, Trinity College, Dublin 2, Ireland}

\emailAdd{bojana.brkic@ff.bg.ac.rs}
\emailAdd{burici@tcd.ie}
\emailAdd{majab@ipb.ac.rs}
\emailAdd{djdusan@ipb.ac.rs}
\emailAdd{latas@ff.bg.ac.rs}

\abstract{Quantum field theory on two- and three-dimensional fuzzy anti-de Sitter spaces is introduced and studied. We find a complete set of solutions to the fuzzy Klein-Gordon equation and identify the commutative limit in which they reduce to classical scalar field modes. After introducing the noncommutative boundary via semi-classical states, the fuzzy modes are used to obtain the boundary two-point function. In both two and three dimensions, this gives an interesting two-parameter deformation of the conformal two-point function. For the fuzzy AdS$_2$, the two-point function is given explicitly in terms of the Appell function $F_1$.}

\begin{document} 
\maketitle

\section{Introduction}

Many attempts to quantise gravity have emerged from theoretical considerations over the past decades. While it is impossible to single out or adopt one approach due to the lack of experimental or observational data, general relativity hints at certain properties that the full quantum theory of gravity should have. In particular, based on the area law of black hole thermodynamics, it has been conjectured that gravity is holographic, meaning that the degrees of freedom related to a spacetime region are encoded at its boundary. The most concrete realisation of this principle through the AdS/CFT correspondence, \cite{Maldacena:1997re,Gubser:1998bc,Witten:1998qj}, has led to numerous investigations of classical and quantum fields on anti-de Sitter space.
\smallskip

The original AdS/CFT correspondence related string theory on AdS$_5\times S^5$ to $\,\mathcal{N}$=4 super Yang-Mills theory in four dimensions, but it has since been extended to other models of gauge/gravity duality, not necessarily involving string theory. Moreover, even without a concrete model of quantum gravity in mind, one can study asymptotically AdS spacetimes with additional matter content and use the well-developed holographic dictionary to extract different properties of the boundary field-theory duals. This procedure, called the \textit{bottom up} approach, has led to a successful holographic description of several systems in condensed matter physics, \cite{Baggioli:2019rrs}. For example, Einstein's gravity with a negative cosmological constant, a $U(1)$ gauge field and a complex scalar in the bulk was shown to lead to superconductor behaviour in the dual field theory, thus realising a holographic superconductor in $(2+1)$ dimensions, \cite{Hartnoll:2008vx}. 
\smallskip

Another aspect of quantum gravity that is common to most approaches, is that at sufficiently small length scales the notion of points becomes meaningless, and the spacetime coordinates come to be described by non-commuting operators. The microscopic origin of noncommutativity varies across different models and will not concern us in this work. Instead, our starting point will be a fixed noncommutative (or \textit{quantum}) spacetime, singled out by symmetry and consistency requirements. From there, one can proceed to define and study quantum field theory on this background, hoping that the difference compared to its commutative counterpart captures the leading quantum-gravitational effects.
\smallskip

Apart from its relation to gravity and models of condensed matter physics, anti-de Sitter space can potentially be used as a regulator for ordinary QFT in flat space. This idea, first proposed in \cite{Callan:1989em}, was recently combined with insights from conformal bootstrap, \cite{Poland:2018epd}, to give a modern take on the S-matrix bootstrap, with promising initial results, \cite{Paulos:2016fap}. Similar questions have also been explored in the noncommutative context. In particular, \cite{Chu:2001xi} studied the potential of the fuzzy sphere, \cite{Madore:1991bw}, to serve as a regulator for quantum field theory on the Moyal-Weyl plane.
\medskip

With the above various roles of AdS spaces in mind, in this work we will initiate the study of quantum field theory on the {\it fuzzy} anti-de Sitter space. For simplicity, we restrict our attention to two and three dimensions. Among various approaches to noncommutative geometry, fuzzy AdS spaces that we use are constructed within the frame formalism of \cite{Madore:2000aq}. The basic idea of the formalism, in short, is the quantisation rule that replaces coordinates {\it and} elements of a moving frame of some classical manifold by operators on a Hilbert space. This procedure is to respect the classical frame relations, which are reproduced by quantum commutators. Along with a number of quantum spaces that have been analysed within this approach, starting with the fuzzy sphere, the formalism has recently been used to formulate noncommutative versions of de~Sitter and anti-de~Sitter spaces in various dimensions, \cite{Buric:2015wta,Buric:2017yes,Buric:2022ton}: we call these the fuzzy (A)dS spaces. One of the benefits of working in the frame formalism is that it allows for a natural definition of classical and quantum field theory, giving a mathematical framework to study  field quantisation and renormalisation on curved noncommutative spaces. The main objectives of this work are, 1) to determine the complete set of free field modes on fuzzy AdS$_2$ and AdS$_3$ spaces, and 2) to compute the corresponding boundary two-point function for the free scalar field. We find that the two-point function reduces to the commutative one in a particular limit of parameters that characterise the space, which is thus identified as the commutative limit.
\medskip

The relation between NC geometry and AdS/CFT is not by itself a new one. There have been several different proposals for noncommutative versions of (A)dS, and within some, dynamics of quantum fields in the bulk and/or on the boundary was studied. In \cite{Jurman:2013ota,Pinzul:2017wch,deAlmeida:2019awj}, the authors proceed by analogy with the fuzzy sphere and define the fuzzy (A)dS$_2$ by promoting the classical embedding coordinates to generators of $SO(2,1)$; the Laplacian is defined as the quadratic Casimir of the symmetry algebra. This approach has the favourable property of keeping the manifest $SO(2,1)$ symmetry. After passing to a Moyal-Weyl description of the product, \cite{Pinzul:2017wch,deAlmeida:2019awj} analyses the boundary two-point function, working to the first order in the $\,\star$-product expansion. This gives an interesting result -- the two-point function turns out to be the same as in the commutative case, up to a rescaling of time. Somewhat differently, $h$- and $q$-deformed versions of the hyperbolic plane were constructed in \cite{Madore:1999fi,Almheiri:2024ayc}. The authors analysed the scalar field two-point function which is largely controlled by the underlying quantum group symmetry and reduces to the commutative case upon setting $h=0$ or $q=1$, respectively. Deformation-based models have also been analysed in dimensions higher than two, including a non-formal deformation quantisation of BTZ black holes (locally isometric to AdS$_3$) carried out in \cite{Bieliavsky:2002ki,Bieliavsky:2004yp}. This analysis was restricted to geometry and did not include quantum fields.
\smallskip

Perhaps the main difference of ours compared to the above approaches lies in the use of semi-classical states, which provide us with notions of classical geometry such as points, regions and the spacetime boundary. In addition, spacetime coordinates, scalar fields etc. are operators on a Hilbert space rather than elements of an abstract algebra. The boundary-two point function for the fuzzy AdS$_2$ computed in this work, \eqref{intro-2pt}, is again a deformation of the commutative one, inequivalent to any of the above given. Our models are perhaps most similar to covariant quantum spaces, see \cite{Sperling:2018xrm,Sperling:2019xar,Karczmarek:2022ejn} and references therein, that arise as solutions to IKKT-like matrix models.\footnote{A closer comparison between our fuzzy dS$_4$ space and that of \cite{Sperling:2018xrm,Sperling:2019xar} was carried out in \cite{Brkic:2024sud}, revealing some differences in the definition of coordinates and momenta, as well as the Laplacian.} It is an interesting open question to try an realise fuzzy AdS spaces in the same manner. 

\paragraph{Summary of results} Let us now describe the results of the present work in more details. We begin in Section \ref{S:Fuzzy AdS bulk in two and three dimensions} by defining fuzzy AdS$_{d+1}$ spaces, with the focus on two and three dimensions ($d=1,2$). Our approach uses several well established constructions, however put together in a very particular way. As already mentioned, the spacetime coordinates are defined as operators acting in a unitary irreducible representation of the symmetry group $SO(d,2)$. They are accompanied by momenta, another set of operators, so as to form a noncommutative frame in the sense of \cite{Madore:2000aq}. The commutation relations in the algebra are used to define differential and tensor calculi, which in turn suffice for the definition of classical field theory. While one of the advantages of the frame formalism is a natural treatment of gauge fields, we will focus on scalar field theory. In the two-dimensional case, the above part of our construction mostly coincides with the $h$-deformed Lobachevsky plane, \cite{Madore:1999fi,Madore:2000aq}, upon Wick-rotating to Lorentzian signature. However, in contrast to \cite{Madore:1999fi,Madore:2000aq}, we will insist that coordinates and momenta are not elements of an abstract algebra, but operators in a certain discrete series representation of $SO(d,2)$, labelled by a negative integer $l$. Therefore, the fuzzy space will be specified by two parameters, $l$ and the constant of noncommutativity $\kbar$. An immediate further observation is that the fuzzy AdS$_3$ comes with more degrees of freedom than its commutative counterpart, which organise into an {\it internal} space $S^1$. The consistency of this fact with the existence of a commutative limit is discussed in detail in later sections.
\smallskip 

The second principal ingredient that we make use of are the so-called semi-classical states. These are defined similarly to generalised coherent states in the sense of Perelomov, \cite{Perelomov:1986uhd}, i.e. as translates of a certain reference state by group elements. In our case, the reference state is the lowest-weight vector of the discrete series representation and translations are performed by means of (exponentials of) momenta. Semi-classical states are used to define the notion of local measurements and therefore make contact with ordinary, commutative geometry.\footnote{In contrast to some approaches, we do not use the semi-classical states to define a quantisation map.}
\smallskip

Sections \ref{S:Scalar field theory on the fuzzy AdS2} and \ref{S:Scalar field on the fuzzy AdS3} are dedicated to the analysis of solutions to the Klein-Gordon equation on the fuzzy AdS$_2$ and AdS$_3$, respectively. We use the method advocated in \cite{Brkic:2024sud}, that can be applied to any noncommutative space of Lie algebra type and reduces the eigenvalue problem of the fuzzy Laplacian, originally an operator equation, to a partial differential equation in ordinary, commutative variables. For AdS spaces, the resulting differential equation is completely integrable and it is straightforward to write down a complete set of eigenfunctions, that is, eigenoperators. In order to understand properties of these solutions, we will study the expectation values of the eigenoperators (i.e., field modes) in semi-classical states. These can be computed exactly, see equations \eqref{NC-modes-AdS2} and \eqref{exact-mode-ads3} for AdS$_2$ and AdS$_3$, respectively. Remarkably, in the limit $|l|\gg\kbar^{-1}\gg1$, these expressions reduce precisely to the commutative field modes. This fact will motivate us to identify the latter as the classical limit. In three dimensions, the fuzzy modes carry three quantum numbers, one more than the commutative ones. However, in semi-classical states, the dependence on the additional quantum number trivialises and one-parameter families of noncommutative solutions behave as the same classical solution.
\smallskip

The complete set of field modes paves the way towards the definition and analysis of quantum field theory. In Section \ref{S:Two-point function and holography}, we consider the simplest such theory, namely the free massive scalar. We shall follow the standard rules of quantisation and demand that creation and annihilation operators corresponding to field modes satisfy canonical commutation relations, thus generating the usual Fock space. After defining correlation functions and establishing the Wick theorem, we proceed to the main subject of the section -- the computation of the boundary two-point function. The latter is computed using the modes of Sections \ref{S:Scalar field theory on the fuzzy AdS2} and \ref{S:Scalar field on the fuzzy AdS3}, and we again write its expectation values in semi-classical states. In two dimensions, the result can be written exactly in terms of standard special functions,
\begin{equation}\label{intro-2pt}
    G_{\partial\partial} = \frac{2^{-1-2\DDelta}\Gamma(\DDelta-2l)^2}{\Gamma(-2l)^2 \Gamma\left(\DDelta+\frac12\right)^2\DDelta} \left(\frac{1}{-l\kbar}\right)^{2\DDelta}\, F_1\left(2\DDelta,\DDelta-2l,\DDelta-2l,2\DDelta+1;\frac{i\langle \hat t_1\rangle}{l\kbar},\frac{-i\langle\hat t_2\rangle}{l\kbar}\right)\ .
\end{equation}
Here, $\DDelta$ is the conformal dimension of the fundamental field $\phi$, related to its mass in the usual way, $\langle\hat t_i\rangle$ are expectation values of the time coordinate in aforementioned semi-classical states and $F_1$ denotes the Appell function. We show that in the limit $|l|\gg\kbar^{-1}\gg1$, \eqref{intro-2pt} reduces to the usual conformal two-point function. In three dimensions, we do not have a closed form expression for the two-point function such as \eqref{intro-2pt}, but can write an expansion in $|l|^{-1}$ and $\kbar$ for it, with all terms given by linear combinations of Appell functions. It is shown that the three-dimensional two-point function also reduces to the conformal two-point function in the commutative limit. We end Section \ref{S:Two-point function and holography} by showing pictorially how the commutative and noncommutative two-point functions compare for certain finite values of $l$ and $\kbar$.
\smallskip

We believe that the above results are a promising start and warrant further investigations of quantum field theory on fuzzy AdS spaces: some of the future directions are discussed in the concluding section. Among these is the analysis of symmetries. Namely, while our construction of fuzzy AdS spaces is heavily based on representation theory of the group $SO(d,2)$, the latter is not a symmetry of these spaces. This is also reflected in the boundary correlator \eqref{intro-2pt}, which does not take the form of a conformal two-point function. A possible CFT interpretation of \eqref{intro-2pt} as a two-point function in the presence of a defect is briefly mentioned. Other topics discussed in the last section include extended families of localised states and their relation to internal degrees of freedom, the inclusion of interactions and extensions to higher dimensions and/or spaces of positive curvature.

\section{Fuzzy AdS bulk in two and three dimensions}
\label{S:Fuzzy AdS bulk in two and three dimensions}

In this section, we review the material necessary for the construction of scalar field theories on the fuzzy AdS$_2$ and AdS$_3$ spaces. The first subsection is dedicated to the ordinary, commutative, anti-de Sitter space. We will write down the Laplacian and its eigenfunctions, before discussing the scalar field propagator and the boundary limit. In the second and third subsection, fuzzy AdS$_2$ and AdS$_3$ spaces are defined. We introduce coordinates, vector fields, differential calculus, including the Laplace-Beltrami operator, and semi-classical states. These are the essential ingredients used to construct noncommutative field modes and two-point functions in later sections.

\subsection{Scalar fields on AdS}
\label{SS:Scalar fields on AdS and holography}

We will mostly study the AdS$_{d+1}$ space using Poincar\'e coordinates $(\tz,\tx^\mu)$ in which the metric reads
\begin{equation}
    \de s^2 = \ell_{\text{\tiny{AdS}}}^2\, \frac{\diff\tz^2 + \eta_{\mu\nu} \diff\tx^\mu \diff\tx^\nu}{\tz^2}\,, \qquad \mu,\nu = 0,\dots,d-1\ .
\end{equation}
Throughout this work, we shall set the AdS radius to $\ell_{\text{\tiny{AdS}}}=1$. The Minkowski boundary sits at $\{\tz=0\}$ and we will write coordinates on it as $(\tx^\mu) = (\t,\tx^i)$. The metric $\eta_{\mu\nu}$ is of mostly-positive signature.
\smallskip

Our main object of study is quantum field theory of a real scalar on the fixed AdS background. The field, denoted $\phi$, is expanded in modes of the free theory. The latter satisfy the Klein-Gordon equation
\begin{equation}
    \nabla^2 \phi = M^2 \phi \equiv \mathbb{\Delta}(\mathbb{\Delta}-d)\,\phi\,, \qquad \mathbb{\Delta} = \frac{d}{2} + \sqrt{M^2 + \frac{d^2}{4}} \equiv \frac{d}{2} + \nu\ .
\end{equation}
We have denoted the mass of the field by $M$ and the conformal dimension of the corresponding field on the boundary by $\DDelta$. For future convenience, we have also defined the quantity $\nu$ in the second equation. In Poincar\'e coordinates, the Laplacian takes the form
\begin{equation}
    \nabla^2 = \tz^2 \left(\partial_{\tz}^2 - \partial_{\t}^2 + \partial_{\tx^i}\partial_{\tx_i}\right) - (d-1)\,\tz\partial_{\tz}\ .
\end{equation}
It commutes with translations in $\t$ and $\tx^i$, leading to eigenfunctions
\begin{equation}\label{commutative-modes}
    f(\tz,\t,\tx^i) = e^{-i(\omega\t - k_i \tx^i)} \tz^{\frac{d}{2}}\left(c_1 J_\nu\left(\sqrt{\omega^2-k^i k_i}\, \tz\right) + c_2 Y_\nu\left(\sqrt{\omega^2-k^i k_i}\, \tz\right) \right)\ .
\end{equation}
The two linearly independent solutions behave near the boundary as $\tz^\DDelta$ and $\tz^{d-\DDelta}$, respectively. For most of this work, we will use the first solution, and denote it by $\phi_{\omega,k}$. The normalisation constant $c_1$ in two and three dimensions is given in Appendix \ref{A:Classical propagator as a sum over modes}.
\smallskip

The propagator on AdS$_{d+1}$ reads, see e.g. \cite{Penedones:2016voo},
\begin{equation}\label{classical-bulk-propagator}
    G_{bb}^{\text{comm}}(\tz_1,\tx_1;\tz_2,\tx_2) = \frac{C_\DDelta}{\zeta^\DDelta}\ _2F_1\left(\DDelta,\DDelta-\frac{d-1}{2};2\DDelta-d+1;-\frac{4}{\zeta}\right)\,,
\end{equation}
where $\zeta$ is the chordal distance between the pair of points $(\tz_1,\tx_1)$ and $(\tz_2,\tx_2)$. In Poincar\'e coordinates,
\begin{equation}
    \zeta=\frac{\eta_{\mu\nu} (\tx_1-\tx_2)^\mu (\tx_1-\tx_2)^\nu +(\tz_1-\tz_2)^2}{\tz_1 \tz_2}\ ,
\end{equation}
and the normalisation constant $C_{\mathbb{\Delta}}$ is given by
\begin{equation}\label{Joao-2pt-normalisation}
    C_{\mathbb{\Delta}} = \frac{\Gamma(\mathbb{\Delta})}{2\pi^{\frac{d}{2}}\, \Gamma\left(\mathbb{\Delta}-\frac{d}{2}+1\right)}\ .
\end{equation}
In Appendix \ref{A:Classical propagator as a sum over modes} we review how the propagator \eqref{classical-bulk-propagator} arises as an integral over the field modes \eqref{commutative-modes}. In this work, we will be particularly interested in correlation functions on the boundary. In the boundary limit the propagator reduces to the (prefactor times the) conformal two-point function
\begin{equation}\label{classical-boundary-two-pt-function}
     G_{bb}^{\text{comm}}(\tz_1,\tx_1;\tz_2,\tx_2) \to (\tz_1\tz_2)^{\DDelta}\, \frac{C_\DDelta}{(\tx_1-\tx_2)^{2\DDelta}} \equiv (\tz_1\tz_2)^{\DDelta}\,  G^{\text{comm}}_{\partial\partial}(\tx_1,\tx_2)\ .
\end{equation}
The last equality defines the two-point function $G^{\text{comm}}_{\partial\partial}(\tx_1,\tx_2)$, which, notice, we take {\it not} to be unit-normalised. This way of obtaining correlation functions of the boundary theory is in line with the extrapolate dictionary, \cite{Harlow:2011ke}. In the commutative case, it gives the same form of the two-point function of operators dual to the scalar field as the standard Gubser-Klebanov-Polyakov-Witten dictionary, \cite{Gubser:1998bc, Witten:1998qj}, but it does not require computations using the (Euclidean) action. 

\subsection{Fuzzy AdS$_2$}

Classically, gravity can be described as geometry of a Riemann or a Riemann-Cartan space. In the tetrad formulation, which is for many purposes in physics more fundamental, the basic field is the {\it moving frame} $\te_\alpha=\te_\alpha^\mu(\tx) \partial_\mu$, where $\{\tx^\mu\}$ denote coordinates, $\alpha=0,\dots,$ are frame indices and $\mu=0,\dots,d$ are coordinate indices. The metric is locally flat, that is in frame indices $\, \tg_{\alpha\beta}=\eta_{\alpha \beta}$. In coordinate indices, $\ \tg_{\mu\nu}(\tx)=\te^\alpha_\mu(\tx) \,\te^\beta_\nu(\tx)\, \eta_{\alpha \beta}\,$, where $\te^\alpha_\mu(\tx)$ is the matrix inverse of the frame $\te_\alpha^\mu(\tx)$. Other characteristics of a Riemann-Cartan space, like connection and torsion can be introduced.

The basic assumption of the NC frame formalism is that quantisation of a curved space is done in a locally flat frame, that is, that vector fields $\te_\alpha$ are quantised as noncommutative derivations $\,e_\alpha =[\hp_\alpha,\ \ ]$. This corresponds to the following quantisation prescription
\begin{equation}
    \te_\alpha^\mu(\tx) = \te_\alpha\tx^\mu \ \mapsto \  e_\alpha^\mu(\hx)=[\hp_\alpha,\hx^\mu]\ ,
\end{equation}
and introduces momenta $\hp_\alpha$ as a noncommutative version of the frame. Relations in the momentum algebra mostly determine differential geometry of a specific NC space: for example, if momenta form a Lie algebra, the corresponding noncommutative space is of constant curvature (assuming, in addition, a torsionless connection), \cite{Madore:2000aq}.

We will not discuss further properties of the NC frame formalism, nor shall we explain the details of the construction of fuzzy spaces that we use: the fuzzy AdS$_3$ was introduced in \cite{Buric:2022ton}, while fuzzy AdS$_2$ is defined in complete analogy with fuzzy dS$_2$, \cite{Brkic:2024sud}.

\smallskip
 The simplest way to quantise AdS$_2$  space is in Poincar\'e coordinates. We use  the following quantisation of coordinates and momenta:
\begin{align}\label{fAdS2-coords-momenta}
  &   \hat z = i\kbar E_+\,, \qquad \hp_z = H\,,\\[4pt]
   &  \hat t =-i \kbar H\,, \qquad \hp_t =E_+\ ,
\end{align}
where $H,E_\pm$ are generators of $SL(2,\mathbb{R})$, acting in a certain discrete series representation $T_l^-$, with $l$ a negative integer or half-integer. Our conventions regarding the group $SL(2,\mathbb{R})$ and its representations are collected in Appendix \ref{A:Representations}. Using the realisation of the discrete series written there, the coordinates and momenta assume the form of differential operators
\begin{align}\label{fAdS2-coords-momenta-in-rep}
    & \hat z = \kbar \xi\,, \hskip4cm \hp_z = \xi \partial_\xi + l+1\,, \\[4pt]
    & \hat t =-i \kbar \left( \xi\partial_\xi+l+1\right)\,, \hskip1.59cm \hp_t = - i \xi\,,
\end{align}
that act on functions $f(\xi)$ of a positive real variable $\xi\,$. Finally, $\kbar$ is a constant, referred to as the constant of noncommutativity.
\smallskip

Differential and pseudo-Riemannian geometry of the fuzzy AdS$_2$ is encoded in coordinate-momentum and momentum-momentum commutation relations. The former, i.e. the frame relations
\begin{equation}\label{frame-relations}
    [\hp_z,\hat z] = \hat z\,, \qquad [\hat p_t, \hat t] = \hat z\,,
\end{equation}
mean that the momenta $\{\hp_z,\hp_t\}$ are to be thought of as noncommutative analogues of the classical vector fields
\begin{equation}\label{classical-frame}
    \te_z = \tz \partial_{\tz}\,, \qquad \te_t = \tz \partial_{\t}\ .
\end{equation}
In the following, we will study eigenfunctions of the scalar Laplace-Beltrami operator, which reads
\begin{equation}\label{Laplacian-fAdS2-abstract}
    \Delta \phi= [\hp_z,[\hp_z,\phi]] - [\hp_z,\phi] - [\hp_t,[\hp_t,\phi]]\ .
\end{equation}
The Laplacian \eqref{Laplacian-fAdS2-abstract} is uniquely specified by the rules of frame formalism. Let us, however, mention that some other versions of the Laplace operator have also been considered in the literature. A common choice that was studied in related models, \cite{Sperling:2018xrm,Brkic:2021lre}, is to drop the linear term $[\hp_z,\phi]$ from \eqref{Laplacian-fAdS2-abstract}. We will call the resulting operator the {\it d'Alembertian} and denote it by $\Box\,$. A comparison of the Laplacian and the d'Alembertian in the case of the fuzzy de Sitter spaces was done in \cite{Brkic:2024sud}, where it was found that their eigenfunctions are related by multiplication by a simple factor. Also, $\Delta$ differs from the quadratic Casimir of $\mathfrak{so}(2,1)$ that was used e.g. in \cite{Jurman:2013ota,Pinzul:2017wch,deAlmeida:2019awj} -- in fact, its centraliser in $SO(2,1)$ is trivial. For reasons to be explained in subsequent sections, we shall use the Laplacian \eqref{Laplacian-fAdS2-abstract} for the remainder of this work.
\smallskip

Note that our definition of the fuzzy AdS$_2$ involves two parameters, $\kbar$ and $l$. The latter characterises the discrete series representation $T_l^-$. Looking ahead, the classical AdS space arises upon taking $|l|$ to be large and $\kbar$ to be small. This will be elaborated in great detail in Sections \ref{S:Scalar field theory on the fuzzy AdS2} and \ref{S:Two-point function and holography}.
\medskip

The notion of local measurements on the fuzzy AdS$_2$ is established through {\it semi-classical states}. The latter are states $|\tx\rangle$ in the Hilbert space, labelled by points of the classical spacetime, such that for a local observable $\mathcal{O}$ with the noncommutative analogue $\mathcal{\hat O}$, the value $\mathcal{O}(\tx)$ is approximated by $\langle \tx|\mathcal{\hat O}|\tx\rangle$. In the case at hand, the semi-classical states are generated by acting with (exponentials of) momenta on the lowest-weight vector $|\Psi_0\rangle\,$ in the Hilbert space, see Appendix \ref{A:Representations},
\begin{equation}\label{semi-classical-states-AdS2}
    |\lambda,c\rangle \equiv \lambda^{-\hat p_z}\, e^{-c \hat p_t} \,|\Psi_0\rangle \ .
\end{equation}
The lowest-weight vector is the solution to equations $\tilde{E}_- |\Psi_0\rangle=0 $,  $ \tilde{H}|\Psi_0\rangle=-l|\Psi_0\rangle$. In our realisation of the discrete series, it is given by the function
\begin{equation}
    \Psi_0 (\xi) = \frac{\xi^{-2l-1} e^{-\xi}}{2^{2l+1/2}\sqrt{\pi\Gamma(-2l)}} \equiv N \xi^{-2l-1} e^{-\xi}\ .
\end{equation}
Expectation values of coordinates in the lowest-weight state are
\begin{equation}
    \langle\Psi_0 |\hz| \Psi_0\rangle = -l\kbar\,, \qquad \langle\Psi_0|\hat t|\Psi_0\rangle = 0\ ,
\end{equation}
and therefore this state corresponds to the classical point $(\tz_0,\t_0)=(-l\kbar,0)$. Expectation values of coordinates in other semi-classical states follow by the frame relations \eqref{frame-relations} and the Baker-Campbell-Hausdorff formula and read
\begin{align}\label{z-exp-AdS2}
    & \langle \lambda,c\,|\hat z|\lambda,c\rangle = \lambda \langle\Psi_0| \hat z | \Psi_0\rangle = - l\kbar \lambda\,,\\[2pt]
    & \langle \lambda,c\,|\hat t|\lambda,c\rangle = \langle\Psi_0| \hat t + c\hat z | \Psi_0\rangle = - l\kbar c\ .\label{t-exp-AdS2}
\end{align}
These expectation values are nothing else but the coordinates of the classical point obtained from $(\tz_0,\t_0)$ by a flow along the integral curves of $\te_{\tz}$ and $\te_{\t}$,
\begin{equation}\label{classical-flow-ads2}
    \lambda^{\te_z}\, e^{c \te_t} \cdot (\tz_0,\t_0) = (\lambda \tz_0,\t_0 + c\tz_0) = (- l\kbar \lambda,- l\kbar c)\ .
\end{equation}
In our realisation of the discrete series, the semi-classical states are 
\begin{equation}\label{coherent-states-explict-AdS2}
    \langle\xi |\lambda,c\rangle =N\, \lambda^l\,\xi^{-2l-1}\, e^{-\frac{1-ic}{\lambda}\, \xi} \, , \qquad N =  \left( 2\pi\, 2^{4l}\, \Gamma(-2l)\right)^{-1/2} \ . 
\end{equation}
From here, one can also see that the semi-classical states satisfy the completeness relation
\begin{equation}\label{resolution-of-the-identity}
    \int\limits_{-\infty}^\infty \diff c \int\limits_0^\infty \frac{\diff\lambda}{\lambda}\, |\lambda,c\rangle\langle\lambda,c| \equiv P  = -\frac{4\pi}{2l+1}\, I\ ,
\end{equation}
where $I$ denotes the identity operator. Indeed, we have
\begin{align}
    \langle\xi|P|\xi'\rangle & = N^2 \int\limits_0^\infty \diff \lambda\, \lambda^{2l-1} (\xi\xi')^{-2l-1} e^{-\frac{\xi+\xi'}{\lambda}} \int\limits_{-\infty}^\infty \diff c\, e^{\frac{i(\xi-\xi')}{\lambda}c}\\
    & = 2\pi N^2 \xi^{-4l-2} \delta(\xi-\xi') \int\limits_0^\infty \diff\lambda\, \lambda^{2l} e^{-\frac{2\xi}{\lambda}} = -\frac{2^{-2l+1}}{2l+1} \xi^{-2l-1}\delta(\xi-\xi')\ . \nonumber
\end{align}
The prefactor multiplying the delta function arises due to the fact that states $|\xi\rangle$ are not normalised to $\delta$-function, see Appendix \ref{A:Representations}. Comparing the last equation with the inner product \eqref{inner-product-discrete-series} in the discrete series gives the resolution of the identity \eqref{resolution-of-the-identity}. The overlaps of the semi-classical states are
\begin{equation}
    \langle\lambda,c\,|\lambda',c'\rangle = \left(\frac{\lambda+\lambda' + i(c\lambda' -c'\lambda)}{2\sqrt{\lambda\lambda'}}\right)^{2l}\ .
\end{equation}
These are all the properties of the semi-classical states that we will need in the following sections. For a general discussion similar generalised coherent states in the context of group theory, see \cite{Perelomov:1986uhd}, and \cite{Steinacker:2024unq} more specifically in noncommutative geometry.

\subsection{Fuzzy AdS$_3$}

A model of the fuzzy AdS$_3$ which satisfies the axioms of the frame formalism was proposed in \cite{Buric:2022ton}. As in two dimensions, coordinates and momenta are constructed using generators of the classical isometry group acting in some irreducible representation. The isometry group of AdS$_3$ is locally isomorphic to $SO(2,2)\sim SL(2,\mathbb{R})\times SL(2,\mathbb{R})$ and we denote its generators by $\{H,E_\pm\}$ and $\{\bar H,\bar E_\pm\}$. The bared generators commute with unbared ones and both sets satisfy the usual $\mathfrak{sl}(2,\mathbb{R})$ brackets. Following \cite{Buric:2022ton}\footnote{More precisely, our coordinate generators have the opposite sign compared to \cite{Buric:2022ton}.}, we put
\begin{align}\label{coord-momenta-ads3-1}
    & \hz= -2i\kbar \sqrt{E_+\bar E_+}\,, \hskip5.8cm \hp_z = H + \bar H\,, \\[4pt]
    & \hx= i\kbar \left( \sqrt{\dfrac{\bar E_+}{E_+}}\, \left(H-\dfrac14\right) + \sqrt{\dfrac{E_+}{\bar E_+}}\, \left(\bar H- \dfrac14\right)\right)\,, \qquad \hp_x = E_+ + \bar E_+\,, \\
    & \htt= i\kbar \left(\sqrt{\dfrac{\bar E_+}{E_+}}\, \left(H-\dfrac14\right) - \sqrt{\dfrac{E_+}{\bar E_+}}\, \left(\bar H-\dfrac14\right)\right)\,, \qquad \hp_t = E_+ - \bar E_+\ .\label{coord-momenta-ads3-3}
\end{align}
The generators act as operators in the representation $T_l^-\otimes \bar T_l^-$ of $\mathfrak{so}(2,2)$. In particular, properties of the discrete series representations ensure that $E_+,\bar E_+$ are positive operators and thus their positive square roots may be defined. Explicitly, using the realisation of the discrete series detailed in Appendix \ref{A:Representations}, we substitute in the above expressions
\begin{equation}
    H = \xi\partial_\xi + l + 1\,, \quad E_+ = -i\xi\,, \qquad \bar H = \bar\xi\partial_{\bar\xi} + l + 1\,, \quad \bar E_+ = -i\bar\xi\,,
\end{equation}
to turn coordinates and momenta into differential operators that act on functions $f(\xi,\bar\xi)$. More generally, we could have allowed for different parameters $l$ and $\bar l$ for the discrete series representations of the two $SL(2,\mathbb{R})$ factors. In the following we will stick to the simplest choice $l=\bar l$, made above.
\smallskip

The discussion of differential geometry and semi-classical states from the previous subsection carries to the present case in the obvious way. The action of the Laplacian on scalars reads
\begin{equation}\label{Laplacian-fAdS3-abstract}
    \Delta \phi = [\hp_z,[\hp_z,\phi]] - 2 [\hp_z,\phi] - [\hp_t,[\hp_t,\phi]] + [\hp_x,[\hp_x,\phi]]\ .
\end{equation}
As in two dimensions, the Laplacian breaks AdS symmetry -- the centraliser of $\Delta$ in $SO(2,2)$ is the one-dimensional subgroup generated by $H-\bar H$. The semi-classical states are given by
\begin{equation}\label{semi-classical-states-AdS3}
 |\lambda,b,c\rangle=\lambda^{-\hat{p}_z}e^{-b\hat{p}_x}e^{-c\hat{p}_t} |\Psi_0\otimes\bar\Psi_0\rangle\,,
\end{equation}
where $|\Psi_0\otimes\bar\Psi_0\rangle $ is the lowest-weight vector in the representation space. Explicitly, as functions of $\xi$ and $\bar\xi$, these states are
\begin{equation}\label{coherent-states-ads3-rep}
   \langle \xi,\bar\xi|\lambda,b,c\rangle= N^2 \lambda^{2l} \left(\xi\bar\xi\right)^{-2l-1} e^{\frac{i}{\lambda}\, \left( \xi(b+c+i)+\bar\xi(b-c+i) \right) }\,  ,
\end{equation}
where the normalisation constant $N$ was introduced in \eqref{coherent-states-explict-AdS2}. 
The overlaps are
\begin{small}
\begin{equation*}
    \langle \lambda,b,c\, |\lambda',b',c'\rangle = \left(\frac{\left(\lambda+\lambda'+i\,((b-c)\lambda'-(b'-c')\lambda)\right) \left(\lambda+\lambda'+i\,((b+c)\lambda'-(b'+c')\lambda)\right)}{4 \lambda \lambda'}\right)^{2l} \ .
\end{equation*}
\end{small}

\vskip10pt

Expectation values of coordinates in semi-classical states are determined as in two dimensions. In the lowest-weight state, we find
\begin{equation*}
    \langle\Psi_0\otimes\bar\Psi_0 |\hz| \Psi_0\otimes\bar\Psi_0\rangle =\frac{\Gamma\left(\frac12-2l\right)^2 }{\Gamma(-2l)^2}\,\kbar\,, \quad \langle\Psi_0\otimes\bar\Psi_0|\hat t|\Psi_0\otimes\bar\Psi_0\rangle = 0=\langle\Psi_0\otimes\bar\Psi_0|\hat x|\Psi_0\otimes\bar\Psi_0\rangle \ .
\end{equation*}

From here, the frame relations lead to expectation values in general states
\begin{align}\label{coord-expectation-ads3-1}
     & \langle \lambda,b,c\,|\hat z|\lambda,b,c\rangle = \frac{\Gamma\left(\frac12-2l\right)^2 }{\Gamma(-2l)^2}\,\kbar\lambda\, ,\\[4pt]
     & \langle \lambda,b,c\,|\hat t|\lambda,b,c\rangle = \frac{\Gamma\left(\frac12-2l\right)^2 }{\Gamma(-2l)^2}\,\kbar c\,, 
     \\[4pt]
   & \langle \lambda,b,c\,|\hat x|\lambda,b,c\rangle = \frac{\Gamma\left(\frac12-2l\right)^2}{\Gamma(-2l)^2}\,  \kbar b\ . \label{coord-expectation-ads3-2}
\end{align}
We will simply write $\langle\hat z\rangle$, $\langle\hat t\rangle$, $\langle\hat x\rangle$, if the semi-classical state in question is understood. In the next sections, we will use the asymptotic form of these expressions for large $|l|$, 
\begin{equation} \label{coordinates.limit}
    \langle \lambda,b,c\,|\hat z|\lambda,b,c\rangle \sim -2 l \kbar\lambda\,, \quad
     \langle \lambda,b,c\,|\hat t|\lambda,b,c\rangle \sim -2 l \kbar c\,, \quad
      \langle \lambda,b,c\,|\hat x|\lambda,b,c\rangle \sim -2 l \kbar b \ .
\end{equation}
\vskip4pt

The main difference between two- and three-dimensional fuzzy AdS spaces is that the latter contains {\it internal} degrees of freedom not present in the commutative case. A quick, non-rigorous, way to appreciate this fact is by counting variables that are used to characterise fuzzy functions. For the fuzzy AdS$_2$, the Hilbert space of the discrete series representation $T_l^-$ is made up of single-variable functions $f(\xi)$. Consequently, the operators $\mathcal{\hat O}$ on this space can be thought of as functions of two variables, the operator's matrix elements $\langle\xi|\mathcal{\hat O}|\xi'\rangle$. This matches the dimension of the commutative space. In the case of the fuzzy AdS$_3$, the Hilbert space consists of two-variable functions $f(\xi,\bar\xi)$ and thus the operators are represented by functions of four variables $\langle\xi,\bar\xi| \mathcal{\hat O}|\xi',\bar\xi'\rangle$, one more than the classical dimension. In Sections \ref{S:Scalar field on the fuzzy AdS3} and \ref{S:Two-point function and holography}, it is exhibited how these additional degrees of freedom are consistent with the existence of a commutative limit.

\paragraph{Remark} Similarly as in two dimensions, see eq. \eqref{classical-flow-ads2}, the semi-classical state \eqref{semi-classical-states-AdS3} corresponds to the point obtained by the flow of vector fields
\begin{equation}\label{classical-frame-ads3}
    \te_z = \tz \partial_{\tz}\,, \qquad \te_t = \tz \partial_{\t}\,, \qquad \te_x = \tz \partial_{\tx}\,,
\end{equation}
from the reference point
\begin{equation*}
    (\tz_0,\t_0,\tx_0)=\left(\frac{\Gamma\left(\frac12-2l\right)^2 }{\Gamma(-2l)^2}\,\kbar,0,0\right)\ .
\end{equation*}
Let us stress that this point is different from the point obtained by acting with the generators of $SO(2,2)$, i.e. from
\begin{equation}
    (\tz,\t,\tx) = \lambda^{H + \bar H} e^{b (E_+ - \bar E_+)}  e^{c (E_+ + \bar E_+)} \cdot (\tz_0,\t_0,\tx_0)\ .
\end{equation}
Indeed, the generators of $SO(2,2)$ act  on the commutative AdS$_3$ space as vector fields
\begin{equation}\label{action-so(2,2)-commutative}
    p_z \equiv H + \bar H  = - \tz \partial_{\tz} - \t \partial_{\t} - \tx \partial_{\tx}\,, \quad p_t \equiv E_+ + \bar E_+ = \partial_{\t}\,, \quad p_x \equiv E_+ - \bar E_+ = \partial_{\tx}\ .
\end{equation}
The two sets of vector fields \eqref{classical-frame-ads3} and \eqref{action-so(2,2)-commutative} can be related in a simple way. The former is the pushforward of the latter by the automorphism
\begin{equation}
    \te_z = \Phi^\ast p_z\,, \ \ \te_t = \Phi^\ast p_t\,, \ \ \te_x = \Phi^\ast p_x\,, \qquad\text{where}\qquad \Phi(\tz,\t,\tx) = \left(\frac{-1}{\tz},\frac{-\t}{\tz},\frac{-\tx}{\tz}\right)\ . 
\end{equation}
Therefore, the semi-classical state \eqref{semi-classical-states-AdS3} corresponds to the point of AdS$_3$ obtained as
\begin{equation}
    (\tz,\t,\tx) = \lambda^{\Phi^\ast(H + \bar H)} e^{b \Phi^\ast(E_+ - \bar E_+)}  e^{c \Phi^\ast(E_+ + \bar E_+)} \cdot (\tz_0,\t_0,\tx_0)\ .
\end{equation}

\section{Scalar field theory on the fuzzy AdS$_2$}
\label{S:Scalar field theory on the fuzzy AdS2}

In this section, we will solve the Klein-Gordon equation on the fuzzy AdS$_2$ and analyse properties of the solutions. In the first subsection, a complete set of free field modes is determined. In the second, we find expectation values of these modes in semi-classical states. It is shown that in the 
limit $l\to-\infty$ the properly normalised modes reduce to their commutative counterparts.

\subsection{Solutions to the noncommutative Klein-Gordon equation}
\label{SS:Solutions to NCKG AdS2}

The aim of this subsection is to determine a complete set of eigenfunctions of the fuzzy Laplacian \eqref{Laplacian-fAdS2-abstract}. This requires solving an equation for operator-valued functions, for which no general solution methods are available. In order to proceed, one may try to choose noncommutative coordinates in which the equation simplifies -- such an approach was used for the fuzzy dS$_2$ and dS$_4$ in \cite{Brkic:2024sud} to reduce the fuzzy Klein-Gordon equation to its commutative analogue. In this work, we will follow a different path, also introduced in \cite{Brkic:2024sud}, which allows to write, for any noncommutative space of Lie algebra type, the fuzzy Klein-Gordon equation as a partial differential equation in ordinary variables. In addition, for AdS spaces the Laplace operator admits sufficiently many first integrals, allowing for the complete solution theory.
\medskip

Recall that the algebra of functions on the fuzzy AdS$_2$ is given by
\begin{equation}\label{fuzzy-functions-AdS2}
    \mathcal{A} = \text{End}(\mathcal{H}_l) \cong T_l^- \otimes (T_l^-)^\ast \cong T_l^- \otimes (T_l^-)^B\ .
\end{equation}
In the last step, we used the fact that taking the dual of a discrete series representation is equivalent to applying the Bargmann automorphism, see Appendix \ref{A:Representations}. We shall realise the first factor in \eqref{fuzzy-functions-AdS2} according to Appendix \ref{A:Representations}, by functions of the variable $\xi\equiv\xi_L$; for the second factor, the variable is denoted by $\xi_R$. Putting everything together and using properties \eqref{Bargmann-automorphism} of the Bargmann automorphism, we can think of $\mathcal{A}$ as consisting of two-variable functions $f(\xi_L,\xi_R)$ and write the momenta as differential operators
\begin{equation}
    \text{ad}_{\hp_z} = \xi_L \partial_{\xi_L} + \xi_R \partial_{\xi_R} + 2l + 2\,, \qquad \text{ad}_{\hp_t} = -i \left(\xi_L - \xi_R \right)\ .
\end{equation}
The expression for the Laplacian \eqref{Laplacian-fAdS2-abstract} as a differential operator in variables $\,(\xi_L,\xi_R)$ directly follows. To find its eigenfunctions, we introduce polar coordinates $(\rho,\varphi)$ through
\begin{equation}
    \xi_L = \rho \sin\varphi \,, \quad \xi_R = \rho\cos\varphi\,, \qquad \rho\in(0,\infty)\,,\ \ \varphi\in \left(0,\frac{\pi}{2}\right)\ .
\end{equation}
The range for the angle $\varphi$ comes form the fact that both $\xi_L$ and $\xi_R$ are positive. This fact will come to play an important role in the following. Let us look for eigenfunctions of $\Delta$ of the form $\phi(\rho,\varphi) = \rho^{-2(l+1)}g(\rho,\varphi)$. In terms of $g(\rho,\varphi)$, the Klein-Gordon equation reads
\begin{equation}
    \rho^2\left(\partial_{\rho}^2 + 2\cos\left(\varphi+\frac{\pi}{4}\right)^2 \right) g = M^2 g\ .
\end{equation}
Since the equation does not involve derivatives with respect to $\,\varphi$, we may consider solutions of the form $g(\rho,\varphi) = f(\rho)\,\delta(\varphi-\varphi_0)$. Defining
\begin{equation}
    \beta = \sqrt{2} \cos\left(\varphi_0+\frac{\pi}{4}\right) \,, \qquad \beta\in(-1,1)\,,
\end{equation}
we obtain the equation for $f(\rho)$
\begin{equation}
    \rho^2 \left(\partial_\rho^2 + \beta^2 \right) f = M^2 f\ .
\end{equation}
The solution that is regular at $\rho=0$ is
\begin{equation}
    f(\rho)=c_\beta \,\sqrt{\rho}\, J_{\nu}\left(\beta\rho\right)\, ,
\end{equation}
where $c_\beta$ is a normalisation constant. We shall denote the corresponding field mode by $\phi_\beta$. Plugging back in the prefactors, we get
\begin{equation}\label{eqNCmodes}
     \phi_\beta(\rho,\varphi)= c_\beta\, \delta(\varphi-\varphi_0) \,\rho^{-2l-\frac32}\, J_{\nu}\left(\beta\rho\right)\ .
\end{equation}
The coefficient $c_\beta$ will be fixed in the next subsection by considering the commutative limit. The modes $\{\phi_\beta\}$ form a complete set of solutions to the fuzzy Klein-Gordon equation.

\subsection{Classical limit and boundary behaviour}

The physical and geometric properties of modes $\phi_\beta$ that were found in the previous subsection are obscured by the choice of variables $(\rho,\varphi)$. Had we used another realisation for the discrete series representations, the field modes would have looked very different. The physically relevant properties of $\phi_\beta$ are revealed by computing their matrix elements between semi-classical states. This is carried out in the remainder of the section.
\medskip

We denote the expectation values of modes in semi-classical states by
\begin{equation}
    \langle\phi_\beta\rangle = \left(2^{2l+1}\pi\right)^2 \int\limits_0^\infty \int\limits_0^\infty (\xi_L\xi_R)^{2l+1} \diff\xi_L \,\diff\xi_R\ \phi_\beta(\xi_L,\xi_R) \langle\xi_L | \lambda,c\rangle \langle\xi_R | \lambda,c\rangle^\ast\ .
\end{equation}
By plugging in the solutions \eqref{eqNCmodes} and the explicit form of semi-classical states \eqref{coherent-states-explict-AdS2}, and integrating over the delta function in $\varphi$, the last expression is reduced to
\begin{equation}
    \langle\phi_\beta\rangle = \frac{2\pi \,\lambda^{2l} c_\beta}{\Gamma(-2l)} \int\limits_0^\infty  \diff\rho \,\rho^{-2l-\frac12}\, J_\nu(\beta\rho) \ e^{-\frac{\rho}{\lambda}\,(\sin\varphi_0+\cos\varphi_0)+\frac{ic\rho}{\lambda}\, (\sin\varphi_0-\cos\varphi_0)}\ .
\end{equation}
To evaluate the integral, we use the formula given in \cite{Gradshteyn}, section 6.621,
\begin{equation}\label{GR-integral}
    \int\limits_0^\infty \mathrm{d}x \, x^{\mu-1}e^{-\alpha x}J_{\nu}(\beta x)=\frac{\left(  \frac{\beta}{2\alpha}\right)^\nu \Gamma(\nu+\mu)}{\alpha^\mu \Gamma(\nu+1)}\ _2F_1\left(\frac{\nu+\mu}{2},\frac{\nu+\mu+1}{2};\nu+1;-\frac{\beta^2}{\alpha^2}   \right),
\end{equation}
which is valid for
\begin{equation}\label{conditions}
    \text{Re}(\alpha+i\beta)>0\,, \qquad \text{Re}(\alpha-i\beta)>0\,, \qquad \text{Re}(\mu+\nu)>0\ . 
\end{equation}
Parameters $\beta$ and $\nu$ have the same meaning in \eqref{GR-integral} and in our setup, while the other  are related by
\begin{equation}
    \mu = -2l +\frac12\,, \qquad \alpha = \frac{\sqrt{2-\beta^2} + i c \beta}{\lambda}\ .
\end{equation}
Thus, the conditions \eqref{conditions} are satisfied and we arrive at the expression for the modes
\begin{equation}\label{NC-modes-AdS2}
    \langle\phi_\beta\rangle = \frac{2\pi\, \lambda^{2l} \,c_\beta}{\Gamma(-2l)} \,\frac{\left(\frac{\beta}{2}\right)^\nu \Gamma(\nu-2l+\frac12)}{\alpha^{\nu-2l+1/2}\,\Gamma(\nu+1)}\ _2F_1\left(\frac{\nu-2l+\frac12}{2}\,,\frac{\nu-2l+\frac32}{2}\,;\nu+1;-\frac{\beta^2}{\alpha^2}\right)\ .
\end{equation}
This is one of the main results of the present section. Recall that $\lambda$ and $c$ are proportional to expectation values of coordinates $\langle\hat z\rangle$ and $\langle\hat t\rangle$, see equations \eqref{z-exp-AdS2}-\eqref{t-exp-AdS2}. Upon substituting these expressions in \eqref{NC-modes-AdS2}, the mode is turned into a function of coordinates. We go on to exhibit some properties of this function. Firstly, for small values of $\lambda\sim \langle \hz \rangle$, one obtains $\langle\phi_\beta\rangle\sim\langle \hz \rangle^{\nu+\frac12}=\langle \hz \rangle^{\DDelta}$, which agrees with the behaviour of commutative modes near the boundary, as explained in Section \ref{SS:Scalar fields on AdS and holography}. This observation will be important later when we compute the boundary two-point function of operators dual to the scalar field on the fuzzy AdS. Had we used noncommutative modes proportional to $Y_\nu(\beta \rho)$ in \eqref{eqNCmodes}, we would have obtained $\langle\phi_\beta\rangle\sim\langle \hz \rangle^{d-\DDelta}$, with $d=1$, which again coincides with the commutative behaviour of the other linearly independent solution. 
\smallskip

More importantly, the modes \eqref{NC-modes-AdS2} reduce to the ones on the ordinary AdS$_2$, \eqref{commutative-modes}, by taking the limit $l\to-\infty$. This result will be at the root of our claim that the latter gives the commutative limit of the fuzzy AdS$_2$. To establish it, we begin from the limit formula relating hypergeometric and Bessel functions, \cite{Watson-Bessel}, section 5.7,
\begin{equation}\label{2F1-Bessel-limit}
   J_\nu(\tilde z)=\lim_{a,b\to\infty}\,  \frac{\left(\frac{\tilde z}{2}\right)^\nu}{\Gamma(\nu+1)}\ _2F_1\left(a,b;\nu+1; -\frac{\tilde z^2}{4ab} \right)\ .  
\end{equation}
Here, $\tilde z,\nu$ can be arbitrary complex numbers and $a,b$ tend to infinity through complex values. We put
\begin{equation}
    a = -l + \frac{\nu+\frac12}{2}\,, \qquad b= -l + \frac{\nu+\frac32}{2}\,,
\end{equation}
which implies
\begin{equation}
    \tilde z^2 = 4 \left(-l + \frac{\nu+\frac12}{2}\right) \left(-l + \frac{\nu+\frac32}{2}\right) \frac{\beta^2}{\left(\sqrt{2-\beta^2}+ic\beta\right)^2}\ \frac{\langle\hz\rangle^2}{l^2\kbar^2}\ .
\end{equation}
Therefore, in the $l\to-\infty\,$ limit $a$ and $b$ go to infinity, while $\tilde z$ remains finite,
\begin{equation}
    \tilde z \sim \frac{2\beta \langle\hz\rangle}{\kbar \left(\sqrt{2-\beta^2}-i\frac{\langle\hat t\rangle}{l\kbar}\beta\right)}\sim \frac{2\beta \langle\hz\rangle}{\kbar\sqrt{2-\beta^2}}\ .
\end{equation}
Applying \eqref{2F1-Bessel-limit} to the expression \eqref{NC-modes-AdS2} for the modes and simplifying, we get 
\begin{align}
    \langle\phi_\beta\rangle & \sim 2\pi c_\beta \sqrt{\frac{2\langle\hz\rangle}{\kbar}} \left(\sqrt{2-\beta^2} - i\frac{\beta\langle\hat t\rangle}{l\kbar}\right)^{2l-1/2} J_\nu\left(\frac{2\beta \langle\hz\rangle}{\kbar\sqrt{2-\beta^2}}\right)\nonumber\\
    & \sim 2\pi c_\beta \left(2-\beta^2\right)^{l-\frac14} e^{-i\, \frac{2\beta\langle\hat t\rangle}{\ \kbar\sqrt{2-\beta^2}}} \,\sqrt{\frac{2\langle\hz\rangle}{\kbar}} \ J_\nu\left(\frac{2\beta \langle\hz\rangle}{\kbar\sqrt{2-\beta^2}}\right)\ .
\end{align}
To get to the second line, we used the familiar limit giving the exponential function, $e^x=\lim\limits_{n\to \infty}\left(1+\frac xn\right)^n$. We shall introduce the suggestive notation
\begin{equation}
    \omega = \frac{2\beta}{\kbar \sqrt{2-\beta^2}}\,, \qquad \omega \in \left( \frac{-2}{\kbar}, \frac{2}{\kbar}\right)\ .
\end{equation}
Then  the limiting form of the modes is written as
\begin{equation}\label{limit-of-modes}
    \langle\phi_\beta\rangle\sim \frac{4\pi c_\beta}{\sqrt{\kbar}} \left(\frac{8}{4+\kbar^2\omega^2}\right)^{l-\frac14} \frac{1}{\sqrt{2}}\ e^{-i\omega\langle\hat t\rangle} \sqrt{\langle\hz\rangle}  \ J_\nu\left(\omega \langle\hz\rangle\right)\ .
\end{equation}
In terms of the dependence on $\langle\hz\rangle$ and $\langle\hat t\rangle$, these are precisely the classical modes of Section \ref{SS:Scalar fields on AdS and holography}. Had we used the d’Alembertian modes instead of the Laplacian ones, the result would have lacked the $\sqrt{\langle\hz\rangle}$ factor in front of the Bessel function, and the order of the Bessel function would have been different from $\nu$. This further motivates our choice of Laplacian over d’Alembertian. Matching the overall normalisations instructs us to put
\begin{equation}\label{normalisation-cbeta}
    c_\beta = \frac{\sqrt{\kbar}}{4\pi} \left(\frac{8}{4+\kbar^2\omega^2}\right)^{-l+\frac14}\ .
\end{equation}
The coefficients $c_\beta$ are not fully determined by the above argument because \eqref{limit-of-modes} is valid only in the $l\to -\infty$ limit and therefore the right hand side of \eqref{normalisation-cbeta} may be multiplied by an arbitrary function of $l$ that tends to one in this limit. Equations \eqref{NC-modes-AdS2} and \eqref{limit-of-modes} are the main results of the present section.

\section{Scalar field on the fuzzy AdS$_3$}
\label{S:Scalar field on the fuzzy AdS3}

This section is dedicated to the Klein-Gordon equation on the fuzzy AdS$_3$ and its solutions. In the first subsection, which is largely similar to Section \ref{SS:Solutions to NCKG AdS2}, a complete set of eigenfunctions of the Laplacian is found. The main difference compared to AdS$_2$ is that now noncommutative modes carry one more quantum number than their commutative analogues. Their expectation values in semi-classical states are determined in the second subsection. Since the calculation follows the same pattern as in two dimensions, most of it is relegated to Appendix \ref{A:Field modes on the fuzzy AdS$_3$}. At the end of the second subsection, we show how taking the $l\to-\infty\,$ limit gives rise to classical modes, the additional quantum number entering only through overall normalisation. The normalisation can be fixed by considering the two-point function, which is done in the next section.

\subsection{Solutions to the Klein-Gordon equation}
\label{SS:Solutions to the Klein-Gordon equation}

We follow the same strategy as in two dimensions and, for this reason, we will be brief. Fuzzy functions, i.e. the elements of the operator algebra
\begin{equation}\label{fuzzy-algebra-ads3}
    \mathcal{A} = \text{End}\left(\mathcal{H}_l \otimes \mathcal{\bar H}_l\right)\,,
\end{equation}
are represented by integral kernels $\phi(\xi_L,\xi_R,\bar\xi_L,\bar\xi_R)$,  with $\,\xi_L,\,\xi_R,\,\bar\xi_L,\,\bar\xi_R\in(0,\infty)$. On them, the momenta act as
\begin{align}
    & \text{ad}_{\hp_z} = \xi_L \partial_{\xi_L} + \xi_R \partial_{\xi_R} + \bar\xi_L \partial_{\bar\xi_L} + \bar\xi_R \partial_{\bar\xi_R} + 4l+4\,,\\[4pt]    
    & \text{ad}_{\hp_t} = -i \left(\xi_L - \xi_R - \bar\xi_L + \bar\xi_R \right)\,, \qquad \text{ad}_{\hp_x} = -i \left(\xi_L - \xi_R + \bar\xi_L - \bar\xi_R \right)\ .
\end{align}
From these expressions, the Laplacian \eqref{Laplacian-fAdS3-abstract} is readily computed. The operator greatly simplifies by passing to variables
\begin{equation}
    \xi_L=\chi_L\, e^{\zeta_L}, \quad \bar\xi_L=\chi_L\, e^{-\zeta_L}, \qquad \xi_R=\chi_R\, e^{\zeta_R}, \quad \bar\xi_R=\chi_R\, e^{-\zeta_R}\,,
\end{equation}
where $\chi_L, \chi_R \in (0,\infty)$ and   $\,\zeta_L, \zeta_R \in (-\infty,\infty)$, and further introducing
\begin{equation}\label{45}
    \rho=\sqrt{\chi_L^2+\chi_R^2}\,, \qquad \tan{\varphi}=\frac{\chi_R}{\chi_L}\,,\qquad \zeta_\pm = \frac12\,(\zeta_L \pm\zeta_R)\,,
\end{equation}
with  $\rho\in(0,\infty)$,  $\,\varphi\in(0,\frac{\pi}{2})\,$,  $\,\zeta_\pm \in (-\infty,\infty)$. Acting on functions of the form $\phi = \rho^{-\frac{8l+7}{2}} \,g(\rho,\varphi,\zeta_+,\zeta_-)$, the Klein-Gordon equation reduces to the following equation for the function $\, g$,
\begin{equation}\label{L3d}
    \left(\rho^2\partial_\rho^2 -4\rho^2 \big(1 - \cosh 2\zeta_- \sin 2\varphi\big) - \frac34\right) g = M^2 g\ .
\end{equation}
Similarly as in two dimensions, the Laplacian contains derivatives with respect to the variable $\rho$ only. We may therefore look for solutions of the form $f(\rho)\delta(\varphi-\varphi_0)h(\zeta_+)\delta(\zeta_- - \zeta_{-0})$, with an arbitrary function $h(\zeta_+)$. We further simplify our ansatz by assuming solutions of the form
\begin{equation}\label{3d-ansztz}
     g(\rho,\varphi,\zeta_+,\zeta_-)= f(\rho)\,\delta(\varphi-\varphi_0)\,\delta(\zeta_+ - \zeta_{+0})\,\delta(\zeta_- - \zeta_{-0})\,  ,
\end{equation}
and denote
\begin{equation}\label{constant-gamma}
    \gamma^2 = \cosh 2\zeta_{-0}\,\sin 2\varphi_0 - 1 \ .
\end{equation}
The general solution for the function $f(\rho)$ is then given in terms of Bessel functions. Picking the solution regular at $\rho=0$ and plugging it back into the mode, we arrive at
\begin{equation}\label{modes-ads3-kernels}
    \phi_\gamma = c_\gamma\, \rho^{-4l-3}\, J_\nu(2\gamma\rho)\, \delta(\varphi-\varphi_0)\,\delta(\zeta_+ - \zeta_{+0})\,\delta(\zeta_- - \zeta_{-0})\,,
\end{equation}
where, recall, $\nu = \sqrt{1+M^2}$. We shall in the following assume that $\gamma^2>0$: this additional condition ensures that the argument of the Bessel function $J_\nu\,$ is real, i.e. that the solution \eqref{modes-ads3-kernels} does not increase exponentially for $\rho\to\infty$. The normalisation constant $c_\gamma$ will be discussed below. The mode $\phi_\gamma$ is characterised by three numbers: $\gamma,\,\zeta_{+0},\,\zeta_{-0}$, and thus it would be more properly denoted by $\phi_{\gamma,\zeta_{+0},\zeta_{-0}}$. As anticipated in the introduction to this section, there is one quantum number more than the scalar modes on  commutative AdS$_3$ space have. In the remainder of the text, we shall usually omit the labels $\,\zeta_{+0},\,\zeta_{-0}$ with the understanding that both $\phi_\gamma$ and $c_\gamma$ implicitly depend on them.

\paragraph{Remark} The range of $\varphi$ that was stated above and follows from the properties of discrete series representations is $(0,\frac{\pi}{2})$. The calculations and analysis of this work remain unchanged if this interval is replaced by its closed counterpart $[0,\frac{\pi}{2}]$. It is also possible to impose periodic boundary conditions on the fields by identifying the endpoints of the interval, $0\sim\frac{\pi}{2}$. Indeed, this would preserve the continuity of the Laplacian \eqref{L3d} which only depends on $\sin2\varphi$. With these latter boundary conditions, the solutions \eqref{modes-ads3-kernels} would get replaced by their linear combinations 
\begin{equation}\label{ads3-periodic-modes}
    \tilde \phi_\gamma = \tilde c_\gamma\, \rho^{-4l-3}\, J_\nu(2\gamma\rho)\, \left(\delta(\varphi-\varphi_0) + \delta\left(\varphi - \frac{\pi}{2} + \varphi_0\right) \right)\,\delta(\zeta_+ - \zeta_{+0})\,\delta(\zeta_- - \zeta_{-0})\,,
\end{equation}
that respect the identification $0\sim\frac{\pi}{2}$. In the following, we shall stick with the modes \eqref{modes-ads3-kernels} -- using \eqref{ads3-periodic-modes} would only change the normalisation constant by a factor of two.

\subsection{Properties of solutions: classical limit and boundary behaviour}

In this subsection, we shall study the expectation values of modes \eqref{modes-ads3-kernels} in semi-classical states and in particular, their classical limit. Since the analysis is similar to that in two dimensions, we relegate most of the derivation to Appendix \ref{A:Field modes on the fuzzy AdS$_3$}.
\smallskip

By definition, the expectation value of the mode $\phi_\gamma$ in the semi-classical state $|\lambda,b,c\rangle$ is given by
\begin{align}\label{modes-AdS3-1}
    &\langle\phi_\gamma\rangle = (2^{2l+1}\pi)^4 \int\limits_0^\infty \int\limits_0^\infty  \int\limits_0^\infty  \int\limits_0^\infty  (\xi_L\bar\xi_L \xi_R\bar\xi_R)^{2l+1} \,\diff\xi_L \,\diff\bar\xi_L \,\diff\xi_R \,\diff\bar\xi_R\\
    & \hskip3.2cm \times \phi_\gamma(\xi_L,\bar\xi_L,\xi_R,\bar\xi_R) \langle\xi_L,\bar\xi_L | \lambda,b,c\rangle \langle\xi_R,\bar\xi_R | \lambda,b,c\rangle^\ast\ .\nonumber
\end{align}
The integrals can be performed exactly, see Appendix \ref{A:Field modes on the fuzzy AdS$_3$}, giving the solution
\begin{align}\label{exact-mode-ads3}
    \langle\phi_\gamma\rangle = 2^{8l+5} \pi^4 N^4 \, c_\gamma\, \lambda^{4l}\, \sin(2\varphi_0) &\ \frac{\Gamma(\nu+1-4l)}{\Gamma(\nu+1)} \,\frac{\gamma^\nu}{\alpha^{\nu+1-4l}}\\
    &\times\ _2F_1\left(\frac{\nu+1-4l}{2}\,,\frac{\nu+2-4l}{2}\,;\nu+1\,; -\frac{4\gamma^2}{\alpha^2}   \right)\ . \nonumber
\end{align}
Constants $N$ and $\gamma$ were defined in \eqref{coherent-states-explict-AdS2} and \eqref{constant-gamma}, respectively, and $\alpha$ is now given by equations \eqref{fs-1}-\eqref{fs-3} and \eqref{alpha-const}. The modes \eqref{exact-mode-ads3} are one of the main results of the present section. Their dependence on positions $\langle\hz\rangle$, $\langle\hat t\rangle$, $\langle\hx\rangle$ can be made explicit by writing out $\lambda$ and $\alpha$ in terms of the former. We shall only do this in the classical limit to which we now turn; one can however immediately note that, again, for small $\lambda$ (implying small $\langle \hz\rangle$) we have $ \langle\phi_\gamma\rangle\sim \lambda^{\nu+1}\sim\langle \hz\rangle^\DDelta$, which coincides with the commutative behaviour near the boundary. As in the two-dimensional case, this fact will make it possible to use a standard extrapolate dictionary as in the commutative case to extract boundary correlation functions. 
\smallskip

The classical limit arises by taking $l\to-\infty$. Using properties of hypergeometric functions, given in Appendix \ref{A:Field modes on the fuzzy AdS$_3$}, the expression \eqref{exact-mode-ads3} reduces to
\begin{equation}\label{modes-ads3-classical-limit}
    \langle\phi_\gamma\rangle = 2^{7/2} \pi^2 (-l)^{1/2} \kbar^{-1} c_\gamma \left(f_{\lambda 0}\right)^{4l-1}\sin(2\varphi_0)\, \frac{1}{\sqrt{4\pi}} \, e^{-i(\omega\langle\hat t\rangle -k\langle\hx\rangle)} \, \langle\hz\rangle \ J_\nu\left(\sqrt{\omega^2 - k^2}\,\langle\hz\rangle \right)\ .
\end{equation}
Note that the condition $\gamma^2\geq 0$ ensures $\omega^2\geq k^2$. Remarkably, the dependence of modes on $\langle\hz\rangle$, $\langle\hat t\rangle$, $\langle\hx\rangle$ coincides with the classical field modes on AdS$_3$: the constants $\omega$ and $k$,  expressed in terms of quantum numbers $\varphi_0,\,\zeta_{-,0}\,,\zeta_{+,0}$, are given in \eqref{omega-k-definition}. Notice that, while the fuzzy modes carry three quantum numbers i.e. one more than the commutative modes, their nontrivial dependence on $\langle\hz\rangle$, $\langle\hat t\rangle$, $\langle\hx\rangle$ comes only through $\omega$ and $k$, while the third quantum number enters only through overall normalisation (this is true even before the limit $l\to-\infty$, $\kbar\to0$). Unlike in two dimensions, we will not require the individual modes \eqref{modes-ads3-classical-limit} to reduce to their commutative counterparts \eqref{classical.modes}. Indeed, the object that is required to have the correct classical limit is the full two-point function. Due to internal degree of freedom present in three dimensions, the ratio of commutative and noncommutative modes in the limit should tend to the square root of the internal space volume, i.e. $\sqrt{\pi/2}$. This leads to
\begin{equation}\label{c-gamma}
    c_\gamma = \kbar \, \frac{\, 2^{2 \DDelta -8 l-\frac72} (-l)^{\DDelta -4 l} \Gamma \left(\frac12-2 l\right)^{8l} \Gamma (-2l)^{2-8 l} \left(f_{\lambda 0}\right)^{1-4l}\, }{\pi^3 \Gamma (\DDelta -4 l)\sin2\varphi_0}\ .
\end{equation}
Again, any other function $\,c_\gamma(l)$ that has the same $l\to-\infty\,$ limit  can be used as normalisation.

\section{Two-point function and holography}
\label{S:Two-point function and holography}

\subsection{Quantum field theory}

One of features of the noncommutative frame formalism is that the quantisation of the field theory follows a natural path, closely analogous to the commutative case. For finite matrix geometries, such as the fuzzy sphere, the most common approach to quantisation is via the path integral. Indeed, on these geometries the path integral reduces to an ordinary Lebesgue integral and provides a regularisation for the quantum field theory. The path integral approach is less straightforward to apply for spaces defined via infinite dimensional representations. In this work, we will instead use canonical methods.
\smallskip

Our main object of interest are $n$-point functions of the fundamental scalar field. We will write $\Phi$ to signify the quantum field, in contrast to the classical noncommutative field $\phi$. The vacuum expectation value of any operator $\mathcal{O}$ will be denoted by $\langle\mathcal{O}\rangle \equiv \langle0|\mathcal{O}|0\rangle$. In particular, we wish to compute expectation values of $n$-point correlators in semi-classical states
\begin{equation}\label{n-point-function}
    G_n(\tx_1,\dots,\tx_n) = \langle \tx_1,\dots,\tx_n | \langle\Phi\otimes\dots\otimes\Phi\rangle| \tx_1,\dots,\tx_n\rangle\ .
\end{equation}
Here, $\tx_i$ are points of the classical manifold, $|\tx_i\rangle$ the corresponding semi-classical states and $|\tx_1,\dots,\tx_n\rangle$ their $n$-fold tensor product. The quantity \eqref{n-point-function} is expected to receive corrections compared to the correlation function of the commutative theory, controlled by the scale of noncommutativity.
\smallskip

To define the middle factor in \eqref{n-point-function}, we begin with the mode expansion of the classical noncommutative field $\phi$,
\begin{equation}\label{field-expansion}
    \phi = \iiint \diff\omega\, \diff k \, \diff s \left(\phi_{\omega,k,s} a_{\omega,k,s} + \phi_{\omega,k,s}^\ast a_{\omega,k,s}^\dagger \right)\ .
\end{equation}
We have written the expansion for the fuzzy AdS$_3$, the more complicated of the two cases we are considering. Moreover, we denote the internal label (previously $\varphi_0$) by $s$ in order to indicate how the construction generalises to other backgrounds. The coefficients $a,a^\dagger$ are promoted to operators satisfying the canonical commutation relations,
\begin{equation}\label{canonical-commutation-reltations}
    [a_{\omega,k,s},a^\dagger_{\omega',k',s'}] = \delta(\omega-\omega')\, \delta(k-k') \,\delta(s-s')\,, \qquad [a,a]=[a^\dagger,a^\dagger] = 0\ .
\end{equation}
These relations are the same as in the commutative theory, except that creation and annihilation operators carry the additional internal index $s$. The Hilbert space of the fuzzy QFT is defined as the Fock space generated by $\{a^\dagger_{\omega,k,s}\}$ from the vacuum state $|0\rangle$ obeying
\begin{equation}\label{vacuum}
    a_{\omega,k,s} |0\rangle = 0\ .
\end{equation}
In expressions \eqref{field-expansion}-\eqref{vacuum}, $\omega$ ranges over positive values that it can assume and $k$ over the interval $(-\omega,\omega)$ -- these are direct generalisations of the ranges in the commutative theory. The additional quantum number $s$ goes over all its values.
\medskip

The simplest correlators \eqref{n-point-function} are one-point functions, which are however seen to vanish identically. Our main focus in this section are two-point functions. As in ordinary quantum field theory, they may be expressed as integrals over modes. Denoting all the quantum numbers collectively by $\Lambda = (\omega,k,s)$, we have
\begin{align}
    G_2 = \langle\Phi\otimes\Phi\rangle & = \iint \diff \Lambda \,\diff\Lambda'\, \langle0|(\phi_\Lambda a_\Lambda + \phi^\ast_\Lambda a^\dagger_\Lambda)\otimes(\phi_{\Lambda'} a_{\Lambda'} + \phi^\ast_{\Lambda'} a^\dagger_{\Lambda'})|0\rangle\\
    & = \iint \diff\Lambda \,\diff\Lambda' \,\langle0|\phi_\Lambda a_\Lambda\otimes \phi^\ast_{\Lambda'} a^\dagger_{\Lambda'}|0\rangle = \iint \diff\Lambda\, \diff\Lambda' \,\phi_\Lambda\otimes\phi^\ast_{\Lambda'} \,\langle0| [a_\Lambda, a^\dagger_{\Lambda'}]|0\rangle\ . \nonumber
\end{align}
Due to the canonical commutation relations \eqref{canonical-commutation-reltations}, we arrive at
\begin{equation}\label{2pt-sum-over-modes}
     \langle\Phi\otimes\Phi\rangle = \iiint \diff\omega \,\diff k\, \diff s\,\langle \phi_{\omega,k,s}\otimes\phi^\ast_{\omega,k,s} \rangle\ .
\end{equation}
The resulting two-point function is End$(\mathcal{H})$-valued and may be further evaluated between any pair of states. Taking these to be (tensor products of) semi-classical states defines the correlator \eqref{n-point-function},
\begin{equation}
    G_2(\tx_1,\tx_2) = \iiint \diff \omega\, \diff k \,\diff s\, \langle \tx_1|\phi_{\omega,k,s}|\tx_1\rangle \langle \tx_2|\phi_{\omega,k,s}|\tx_2\rangle^\ast\ .
\end{equation}
Continuing along these lines, we see that the free theory on the fuzzy AdS space satisfies Wick's theorem. Therefore, all $n$-point functions \eqref{n-point-function} are simply obtained from the two-point function.

\subsection{Two-point function on the fuzzy AdS$_2$}
\label{SS:Two-point function on the fuzzy AdS2}

The aim of this subsection is to compute the two-point function of $\Phi$ on the fuzzy AdS$_2$. The two-point function is given by the AdS$_2$ analogue of \eqref{2pt-sum-over-modes},
\begin{equation}\label{2pt-function-AdS2-abstract}
    G_2 = \int \diff \omega\ \phi_\omega \otimes \phi_\omega^\ast\ .
\end{equation}
Geometric properties of $G_2$ are revealed by computing its expectation value in a pair of semi-classical states $|\lambda_i,c_i\rangle$, $i=1,2$. Thus, we consider
\begin{equation}\label{cutoff}
    G_2(\lambda_i,c_i) =  \int \diff \omega\, \langle\lambda_1,c_1|\phi_\beta|\lambda_1,c_1\rangle\, \langle\lambda_2,c_2| \phi^\ast_\beta |\lambda_2,c_2\rangle\ .
\end{equation}
Let us firstly briefly discuss the range of integration. In the commutative case, the frequency assumes values $\omega\in(-\infty,\infty)$ and positive frequency modes have $\omega\in(0,\infty)$. On the fuzzy space, we have $\beta\in(-1,1)$ and therefore $\omega\in(-2\kbar^{-1},2\kbar^{-1})$. The positive frequency modes thus have $\omega\in(0,2\kbar^{-1})$. This is in line with the intuition that the classical space arises when $\kbar$ is small. Substituting the expression \eqref{NC-modes-AdS2} for the modes, we have
\begin{align}\label{G2-AdS2-matrix-elements}
    G_2(\lambda_i,c_i) &= (2\pi)^2\, \frac{\Gamma(\nu-2l+\frac12)^2}{\Gamma(-2l)^2 \Gamma(\nu+1)^2} \,(\lambda_1\lambda_2)^{2l}\\
    & \qquad \times\int\limits_0^{2\kbar^{-1}} \diff \omega\  c_\beta^2 \, \frac{\left(\frac{\beta}{2}\right)^{2\nu}}{(\alpha_1\alpha_2^\ast)^{\nu-2l+1/2}}\,\prod_{i=1}^2\ _2F_1\left(\frac{\nu-2l+\frac12}{2},\frac{\nu-2l+\frac32}{2};\nu+1;-\frac{\beta^2}{\alpha_{(i)}^2}\right)\ .\nonumber 
\end{align}
For compactness of notation, we have put in the last line $\alpha_{(1)} = \alpha_1$ and $\alpha_{(2)}=\alpha_2^\ast$. While this integral looks formidable, it may be evaluated exactly in the boundary limit. In this limit, $\alpha_i\to\infty$, and therefore hypergeometric functions in \eqref{G2-AdS2-matrix-elements} tend to one. We thus get
\begin{equation}
      \tilde G_{\partial\partial}(\lambda_i,c_i) = 4\pi^2 \,\frac{\Gamma(\nu-2l+\frac12)^2}{\Gamma(-2l)^2 \Gamma(\nu+1)^2} (\lambda_1\lambda_2)^{2l} \, \int\limits_0^{2\kbar^{-1}} \diff\omega\ c_\beta^2 \, \frac{\left(\frac{\beta}{2}\right)^{2\nu}}{(\alpha_1\alpha_2^\ast)^{\nu-2l+1/2}}\ .
\end{equation}
We have put a tilde to signify that the boundary propagator requires a further inclusion of the conformal factor, $G_{\partial\partial} = \langle\hat z_1\rangle^{-\Delta} \langle\hat z_2\rangle^{-\Delta} \tilde G_{\partial\partial}$. Expressing everything in terms of expectation values of coordinates,
 \begin{align}
    \tilde G_{\partial\partial} =  &\  4\pi^2 \frac{\Gamma(\nu-2l+\frac12)^2}{\Gamma(-2l)^2 \Gamma(\nu+1)^2} \left(\frac{\langle\hat z_1\rangle}{-l\kbar}\right)^{\nu+1/2} \left(\frac{\langle\hat z_2\rangle}{-l\kbar}\right)^{\nu+1/2}\\
    & \ \times \int\limits_0^{2\kbar^{-1}} \diff \omega\ c_\beta^2\, (2-\beta^2)^{2l-1/2} \left(1-\frac{i\omega\langle\hat t_1\rangle}{2l}\right)^{2l-\nu-1/2} \left(1+\frac{i\omega\langle\hat t_2\rangle}{2l}\right)^{2l-\nu-1/2}\left(\frac{\kbar\omega}{4}\right)^{2\nu} .\nonumber 
\end{align}
We see that the conformal prefactor precisely cancels the dependence on $\langle\hat z_i\rangle$, leaving $G_{\partial\partial}$ as a function of $\langle\hat t_i \rangle$ only. Committing to the choice of normalisation \eqref{normalisation-cbeta}, we arrive at
\begin{equation}\label{2pt-function-single-integral}
    G_{\partial\partial} =  \frac{\Gamma(\DDelta-2l)^2}{4^{2\DDelta}\Gamma(-2l)^2 \Gamma\left(\DDelta+\frac12\right)^2} \left(\frac{1}{-l}\right)^{2\DDelta} \int\limits_0^{2\kbar^{-1}} \diff \omega\ \left(1-\frac{i\omega\langle\hat t_1\rangle}{2l}\right)^{2l-\DDelta} \left(1+\frac{i\omega\langle\hat t_2\rangle}{2l}\right)^{2l-\DDelta}\omega^{2\DDelta-1} \ .
\end{equation}
We have expressed the result in terms of $\DDelta$ rather than $\nu=\DDelta-1/2$, the former being a more appropriate quantum number on the boundary. Notice that the only dependence of the two-point function on $\kbar$ comes through the region of integration. The representation \eqref{2pt-function-single-integral} is expected to have the correct commutative limit. Indeed, by taking $l\to-\infty$, the integrand reduces to $\ \omega^{2\DDelta-1} e^{-i\omega\langle\hat t_{12}\rangle}$, just as in the commutative theory. Further, the $\kbar\to0$ limit makes the region of integration $(0,\infty)$ as is in the commutative case. However, one needs to make sure that the integral over modes converges and commutes with the limits $l\to-\infty$ and $\kbar\to0\,$.
\smallskip
\begin{figure}[h!]
    \centering
    \includegraphics[scale=0.8]{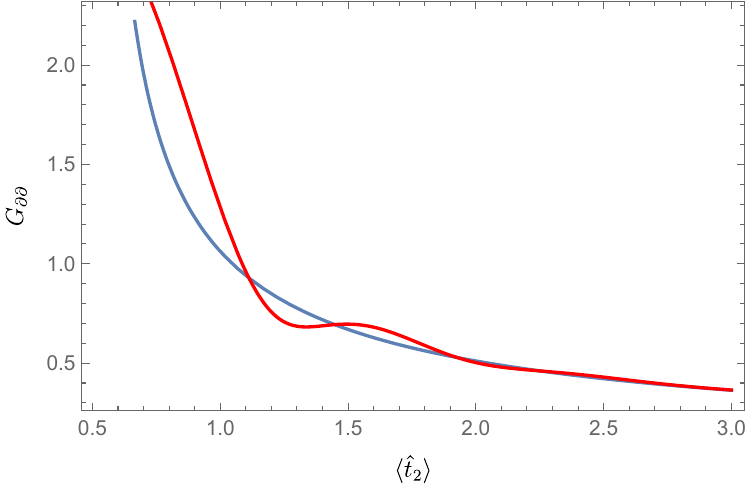}
    \caption{Commutative and the fuzzy two-point function: $l=-30$, $\kbar = 1/4$, $\DDelta=1/3$, $\t_1 = 1/2$}
    \label{plot1}
\end{figure}
Notice that, even in the commutative case, the integral
\begin{equation}\label{commutative-two-point-function}
    G_{\partial\partial}^{\text{comm}} \equiv \frac{4^{-\DDelta}}{\Gamma(\DDelta+\frac12)^2}\int\limits_0^\infty \diff \omega\, \omega^{2\DDelta-1} e^{-i\omega\langle\hat t_{12}\rangle} = C_\DDelta \, \langle i\hat t_{12}\rangle^{-2\DDelta}\,,
\end{equation}
is valid only for $\DDelta<1/2$. We may define $G_{\partial\partial}^{\text{comm}}$ for $\DDelta>1/2$ by the right hand side of \eqref{commutative-two-point-function}, i.e. by analytic continuation in $\DDelta$. This is indeed the usual conformal two-point function, with the normalisation factor $C_\DDelta$ given in \eqref{Joao-2pt-normalisation}. We shall follow the same strategy of analytically continuing in $\DDelta$ to define the two-point function in the noncommutative case. 
\smallskip

\begin{figure}[h!]
    \centering
    \includegraphics[scale=0.8]{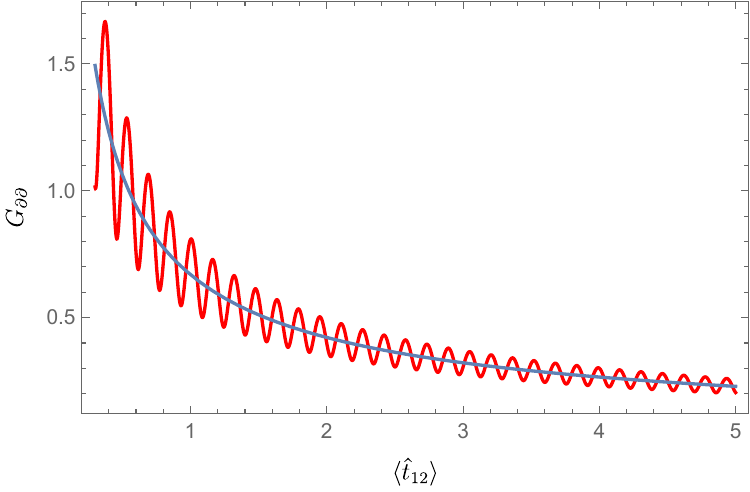}
    \caption{Commutative and the fuzzy two-point function: $l=-\infty$, $\kbar = 1/20$, $\DDelta=1/3$}
    \label{plot2}
\end{figure}

Let us now go back to the integral representation \eqref{2pt-function-single-integral} of the fuzzy two-point function. As it turns out, the integral can be evaluated exactly in terms of the Appell function $F_1$,
\begin{equation}\label{fuzzy-2pt-AdS2-final}
    G_{\partial\partial} = \frac{2^{-1-2\DDelta}\Gamma(\DDelta-2l)^2}{\Gamma(-2l)^2 \Gamma\left(\DDelta+\frac12\right)^2\DDelta} \left(\frac{1}{-l}\right)^{2\DDelta}  \kbar^{-2\DDelta}\, F_1\left(2\DDelta,\DDelta-2l,\DDelta-2l,2\DDelta+1;\frac{i\langle \hat t_1\rangle}{l\kbar},\frac{-i\langle\hat t_2\rangle}{l\kbar}\right)\ .
\end{equation}
This is one of the main results of the present work. Based on the above discussion, \eqref{fuzzy-2pt-AdS2-final} is valid when $\DDelta<1/2$ and should be analytically continued for $\DDelta>1/2$. In the former regime, one can directly take the limit to the commutative two-point function. Firstly, $l\to-\infty$ corresponds to the confluent limit of the Appell function, \cite{Appell}, XLI.I. We obtain
\begin{equation}
    \lim_{l\to-\infty}G_{\partial\partial} = \frac{\kbar^{-2\DDelta}}{2\Gamma(\DDelta+\frac12)^2\,\DDelta}\, _1F_1\left(2\DDelta;2\DDelta+1;\frac{-2i\langle\hat t_{12}\rangle}{\kbar}\right)\ .
\end{equation}
Further taking the limit $\kbar\to0$, using the formula 13.5.1 of \cite{Abramowitz} we obtain 
\begin{equation}
   \lim_{\kbar\to0} \lim_{l\to-\infty} G_{\partial\partial}  =  \frac{\kbar^{-2\DDelta} \Gamma(2\DDelta+1)}{2\Gamma(\DDelta+\frac12)^2\,\DDelta} \left(\frac{2i\langle\hat t_{12}\rangle}{\kbar}\right)^{-2\DDelta} = C_\DDelta\,  \langle i\hat t_{12} \rangle^{-2\DDelta}\ .
\end{equation}
Therefore, we obtain precisely the commutative two-point function given in \eqref{classical-boundary-two-pt-function}.
\smallskip

For finite $l$ and $\kbar$, the fuzzy two-point function \eqref{fuzzy-2pt-AdS2-final} provides a two-parameter deformation of the classical one. In Figures \ref{plot1} and \ref{plot2} we show some examples of how the commutative (blue) and noncommutative (red) results compare for particular values of $l$, $\kbar$ and $\DDelta$. In Figure \ref{plot1} we also fix one of the times to $\t_1=1/2$, which is important because the noncommutative two-point function is not translation invariant. For the second plot, we put $l=-\infty$, so the translation invariance is restored.

\subsection{Two-point function on the fuzzy AdS$_3$}

The last observable that we study in this work is the boundary two-point function on the fuzzy AdS$_3$. On the boundary, the variable $\alpha$ tends to $\infty$, and therefore the mode \eqref{exact-mode-ads3} reduces to 
\begin{align}\label{boundary-mode-ads3}
    \langle\phi_\gamma\rangle = \frac{8\pi^2}{\Gamma(-2l)^2} \,c_\gamma \left(\frac{\Gamma(-2l)^2}{\Gamma(\frac12-2l)^2\kbar}\langle\hz\rangle\right)^{4l} &\sin(2\varphi_0)\, \frac{\Gamma(\nu+1-4l)}{\Gamma(\nu+1)} \ \gamma^\nu\\
    & \times \left(\frac{-4l\kbar f_\lambda^0}{\langle\hz\rangle}\left( 1 + \frac{i\omega\langle\hat t\rangle}{4l} - \frac{ik\langle\hx\rangle}{4l} \right)\right)^{4l-1-\nu}\ .\nonumber
\end{align}
Taking out the conformal factor and committing to the normalisation \eqref{c-gamma} of modes, we get
\begin{equation}
    \frac{\langle\phi_\gamma\rangle}{\langle\hz\rangle^\DDelta} = \frac{2^{\frac12-\DDelta}}{\pi\Gamma(\DDelta)}\left(\omega^2-k^2\right)^{\frac{\nu}{2}} \left(1 - \frac{i\omega\langle\hat t\rangle}{4l} + \frac{ik\langle\hx\rangle}{4l} \right)^{4l-1-\nu}\ .
\end{equation}
The boundary two point-function is therefore
\begin{align*}
    G_{\partial\partial} &= \iiint \diff \omega\, \diff k \, \diff \varphi_0\\
    & \frac{2^{1-2\DDelta}}{\pi^2\Gamma(\DDelta)^2} \left(\omega^2-k^2\right)^{\DDelta-1} \left(1 + \frac{i\omega\langle\hat t_1\rangle}{4l} - \frac{ik\langle\hx_1\rangle}{4l} \right)^{4l-\DDelta} \left(1 - \frac{i\omega\langle\hat t_2\rangle}{4l} + \frac{ik\langle\hx_2\rangle}{4l} \right)^{4l-\DDelta}\ .
\end{align*}
In the $\omega$-$k$ plane, the integration is performed over the triangle defined by points $(0,0)$, $\left(\frac{2}{\kbar},-\frac{2}{\kbar}\right)$ and $\left(\frac{2}{\kbar},\frac{2}{\kbar}\right)$. Indeed, from the definition of $\omega$ and $k$, keeping in mind that $0<\varphi_0<\pi/2$, we have $|\omega|,|k|<2\kbar^{-1}$. The normalisability condition $\gamma^2>0$ can be recast as $\omega^2-k^2\geq 0$. Finally, we set $\omega\geq0$ as we are interested in positive frequency modes. The $(\omega,k)$ integration region is independent of $\varphi_0$, and therefore, the integration over this quantum number is trivial. 
\smallskip

We find it convenient to pass to lightcone coordinates and momenta
\begin{equation}
    k_\pm = \omega \pm k\,, \qquad u^\pm = \frac12\left(\langle\hat t\rangle\pm \langle\hat x\rangle\right)\,,
\end{equation}
in which the relation to AdS$_2$ is most clearly manifested. Putting everything together, the two-point function becomes
\begin{align}\label{ads3-fuzzy-2pt-function}
    G_{\partial\partial} = \frac{2^{-1-2\DDelta}}{\pi\Gamma(\DDelta)^2} \int\limits_0^{4\kbar^{-1}} &\diff k_+\ \left(1 + \frac{i k_+ u^-_1}{4l} \right)^{4l-\DDelta} \left(1 - \frac{i k_+ u^-_2}{4l}\right)^{4l-\DDelta}\\
   \ \times \int\limits_0^{4\kbar^{-1}-k_+} &\diff k_- (k_+ k_-)^{\DDelta-1} \left(1 + \frac{i k_- u^+_1}{4l+ik_+ u_1^-} \right)^{4l-\DDelta} \left(1 - \frac{i k_- u^+_2}{4l-ik_+ u_2^-} \right)^{4l-\DDelta}\nonumber\ .
\end{align}
The integral over $k_-$ is of the type we encountered in two dimensions and produces an Appell $F_1$ function,
\begin{align}
    G_{\partial\partial} = & \frac{2^{-1-2\DDelta}}{\pi\Gamma(\DDelta)^2} \int\limits_0^{4\kbar^{-1}} \diff k_+\, k_+^{\DDelta-1} \left(1 + \frac{i k_+ u^-_1}{4l} \right)^{4l-\DDelta} \left(1 - \frac{i k_+ u^-_2}{4l}\right)^{4l-\DDelta}\\
    &\times\left(4\kbar^{-1} - k_+\right)^{\DDelta} F_1\left(\DDelta,\DDelta-4l,\DDelta-4l,\DDelta+1;\frac{-i\left(4\kbar^{-1} - k_+\right)u_1^+}{4l+ik_+ u_1^-},\frac{i\left(4\kbar^{-1} - k_+\right) u_2^+}{4l-ik_+ u_2^-}\right)\nonumber\ .
\end{align}

\begin{figure}[b!]
    \centering
    \includegraphics[scale=0.75]{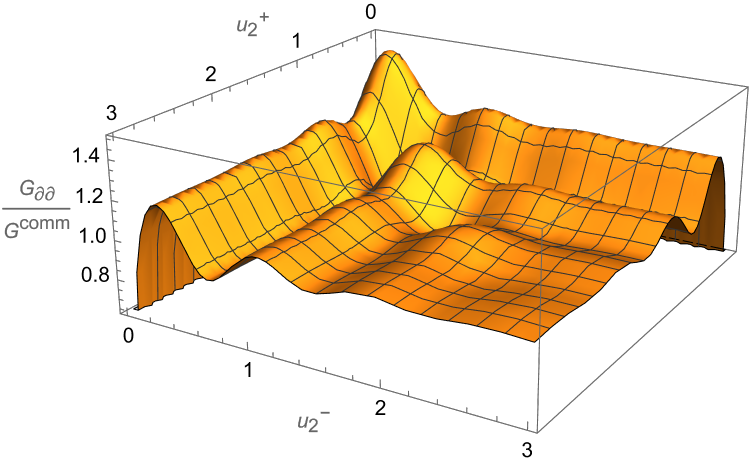}
    \caption{Commutative and the fuzzy two-point function: $l=-20$, $\kbar = 1/2$, $\DDelta=1/3$}
    \label{plot3}
\end{figure}

\begin{figure}[h!]
    \centering
    \begin{minipage}{0.4\textwidth}
         \centering
        \includegraphics[width=0.9\textwidth]{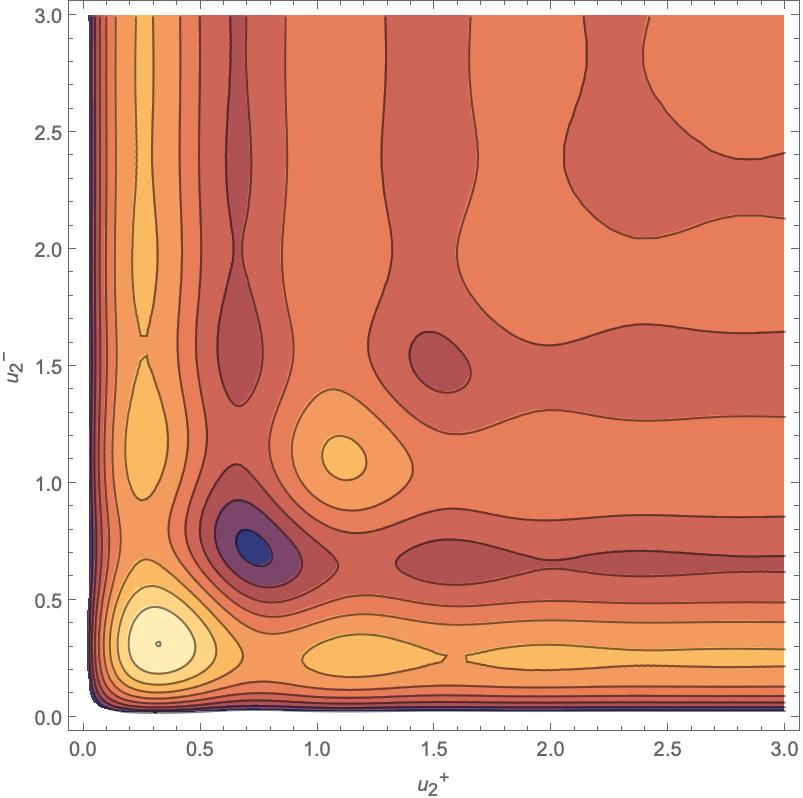}
    \end{minipage}
    \begin{minipage}{0.4\textwidth}
        \centering
        \includegraphics[width=0.9\textwidth]{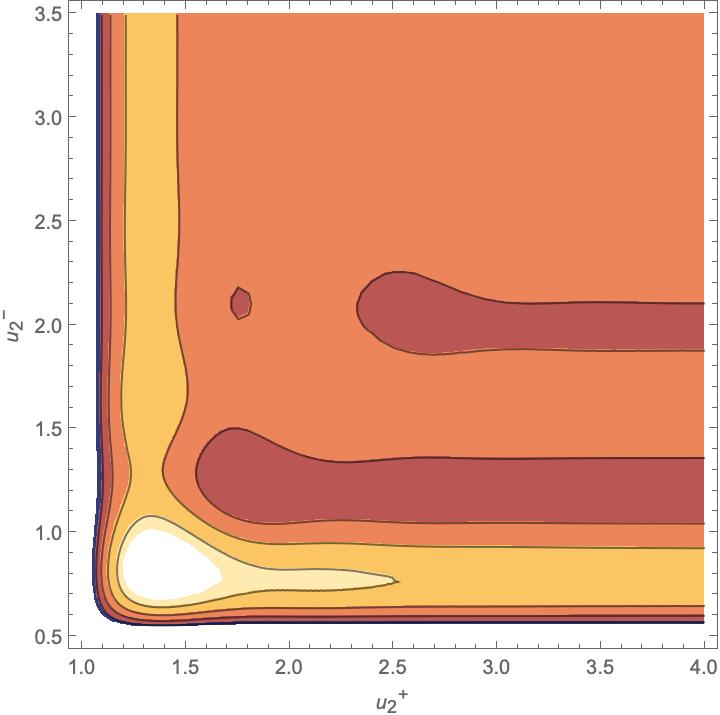}
    \end{minipage}
    \caption{Ratio of noncommutative and commutative two-point functions: $(u_1^+,u_1^-)=(0,0)$ vs $(u_1^+,u_1^-)=(1,1/2)$}
    \label{plot4}
\end{figure}

We do not know if the remaining integral over $k_+$ can be expressed in terms of standard special functions. To proceed, we expand the second line above in powers of $k_+$. Looking at how $k_+$ appears in relation to $\kbar$ and $l$, it is seen that this power series corresponds to that around the classical limit. In the leading order we obtain
\begin{align}\label{2-pt-AdS3-leading-order}
    G^{(0)}_{\partial\partial} =& \frac{2^{-1-2\DDelta}}{\,\pi\Gamma(\DDelta+1)^2} \left(\frac{4}{\kbar}\right)^{2\DDelta} F_1\left(\DDelta,\DDelta-4l ,\DDelta-4l,\DDelta+1;\frac{-iu_1^-}{l\kbar},\frac{iu_2^-}{l\kbar}\right)\\
     & \hskip5.2cm \times F_1\left(\DDelta,\DDelta-4l,\DDelta-4l,\DDelta+1;\frac{-iu_1^+}{l\kbar},\frac{iu_2^+}{l\kbar}\right)\nonumber\ .
\end{align}
Higher orders all give Appell-type integrals and can be computed exactly and straightforwardly. They are suppressed by powers of $\kbar$ and/or $l^{-1}$. Indeed, one can see that the $n$-th order only involves terms that come with powers $\kbar^a (l^{-1})^b$ such that $a+b\geq n$. The expressions are bulky, so we write only the first subleading order:
\begin{align*}
    & G^{(1)}_{\partial\partial} = G^{(0)}_{\partial\partial} + \frac{2^{-3-2\DDelta}}{\pi\Gamma(\DDelta)^2\DDelta} \left(\frac{4}{\kbar}\right)^{2\DDelta} F_1\Big(\DDelta;\DDelta-4 l,\DDelta-4 l;\DDelta+1;\frac{-iu^-_1}{l\kbar},\frac{i u^-_2}{l\kbar}\Big)\\
    & \times\left(-\kbar F_1\Big(\DDelta;\DDelta -4l,\DDelta -4l;\DDelta +1;\frac{-i u^+_1}{l\kbar},\frac{i u^+_2}{l\kbar}\Big)  \right.\\
    & \qquad+ \frac{4l-\DDelta}{(\DDelta+1) l^2 \kbar} \left(u^+_1 (u^-_1 - il\kbar)\, F_1\Big(\DDelta+1;\DDelta-4l+1,\DDelta-4l;\DDelta+2;\frac{-i u^+_1}{l\kbar},\frac{i u^+_2}{l\kbar}\Big) \right.\\
    & \hskip3cm \left.\left. + u^+_2 (u^-_2 +il\kbar) \, F_1\Big(\DDelta+1;\DDelta-4l,\DDelta-4l+1;\DDelta+2;\frac{-i u^+_1}{l\kbar},\frac{i u^+_2}{l\kbar }\Big)\right)\right)\ .
\end{align*}
Let us return to the leading order two-point function \eqref{2-pt-AdS3-leading-order}. The factorised form in lightcone coordinates together with our results for the AdS$_2$ two-point function immediately imply that \eqref{2-pt-AdS3-leading-order} has the correct commutative limit. Indeed, repeating essentially the same steps as above, we find
\begin{align}
   \lim_{\kbar\to0} \ \lim_{l\to-\infty} G^{(0)}_{\partial\partial} & =  \lim_{\kbar\to0}\  \frac{2^{-1-2\DDelta}}{\,\pi\Gamma(\DDelta+1)^2} \left(\frac{4}{\kbar}\right)^{2\DDelta}\  _1F_1\left(\DDelta,\DDelta+1;\frac{4iu_{12}^-}{\kbar}\right)\ _1F_1\left(\DDelta,\DDelta+1;\frac{4iu_{12}^+}{\kbar}\right) \nonumber\\[4pt]
   &  = \frac{2^{-2\DDelta}}{2\pi} \ \left(-iu_{12}^-\right)^{-\DDelta}\left(-iu_{12}^+\right)^{-\DDelta}\ .
\end{align}
Thus, we have arrived at the commutative two-point function
\begin{equation}
    G^{\text{comm}}_{\partial\partial}(\t_1,\tx_1;\t_2,\tx_2) = \frac{1}{2\pi} \,\frac{1}{(-\t_{12}^2 + \tx_{12}^2)^\DDelta} = \frac{1}{2\pi} \,\frac{1}{(-4 \tu^-_{12} \tu^+_{12})^\DDelta}\ .
\end{equation}
Away from the commutative limit, we may plot the fuzzy two-point function \eqref{ads3-fuzzy-2pt-function} and inspect how it differs from its commutative analogue. In Figure \ref{plot3} we show the ratio of the two for particular values of parameters $l$, $\kbar$ and $\DDelta$.\footnote{The plot in Figure \ref{plot3} is obtained from \eqref{ads3-fuzzy-2pt-function} by numerical integration. Alternatively, we may use the expansion explained above and plot $G_{\partial\partial}^{(n)}$ for $n=0,1,2,\dots$. This gives similar results.}. The first point is fixed at the origin, $u_1^\pm=0$. We see that the most significant deviations from the commutative case occur when the second point is also close to the origin. If the first point is not at the origin, Figure \ref{plot4}, the result is, unlike the commutative two-point function, no longer symmetric in $u_{12}^+$ and $u_{12}^-$.

\section{Summary and perspectives}

In this work we initiated the study of quantum field theory on fuzzy AdS spaces. Focusing on two and three dimensions, we determined a complete set of modes for the (noncommutative) free scalar field and computed the boundary two-point function. By analysing expectation values of modes inside {\it semi-classical} states, we identified the commutative limit
\begin{equation}\label{conclusions-commutative-limit}
   |l| \gg\kbar^{-1}\gg1\,,
\end{equation}
in which the fuzzy two-point function reduces to the commutative one. In three dimensions, the derivation exhibits an interesting mechanism by which the additional degrees of freedom present on the noncommutative space are made consistent with the classical limit.
\smallskip

The way how the classical limit arises, even in two dimensions, is technically quite intricate. Let us recap the main points. As mentioned, the classical regime involves taking two limits: the representation label is sent to infinity, $l\to-\infty$, and the noncommutativity parameter to zero, $\kbar\to 0$ (we take from the start that $\ell_{\text{\tiny{AdS}}}=1$). Both these types of limits are familiar in other examples of noncommutative spaces. Usually either only $\kbar$ is present, and treated as a formal parameter, or both labels occur but can be related to one another by a Casimir-type relation. Our case is different from either of the above and there is a curious interplay between the noncommutativity parameter $\kbar$ and the choice of the discrete series representation of $SL(2,\mathbb{R})$. As was explained, the fact that discrete series are realised by functions of a {\it positive} variable $\xi$ led to the restriction on the angle $\varphi\in (0,\frac{\pi}{2})$ and consequently the UV frequency cut-off, $|\omega|<\frac{2}{\kbar}$. Had we used the principal series representation, the cut-off would not have existed. In principle, we are still free to impose a relation between $l$ and $\kbar$, as long as it is consistent with \eqref{conclusions-commutative-limit}. It will be interesting to see if such a relation is implied by consistency/classical limit of some other observables not studied in this work.
\smallskip

While these initial results are very promising, a number of open questions about the status of quantum field theory on fuzzy AdS spaces remain. The main future direction that we have in mind is to go beyond the free theory and study interactions. To this end, the frame formalism provides natural analogues for most concepts of ordinary curved space QFT, with a few omissions. For instance, it appears that our definition of the Laplacian implies that one should not use the trace as the noncommutative analogue of the integral, but modify it by a certain factor.\footnote{The analogous factor in the case of the fuzzy dS$_4$ was written down in \cite{Brkic:2024sud}.} A simplistic approach, at least to boundary correlators of weakly coupled theories, is to write the latter in terms of Witten diagrams as in the commutative case, however with all propagators being fuzzy. E.g. for the $\lambda \phi^3$ theory, the boundary three-point function would be given by
\begin{equation}\label{interactions-simplistic}
    G_3(\tx_1,\tx_2,\tx_3) = \lambda \int\limits_{\text{AdS}} \diff\tz\,\diff^d\tx\,\sqrt{|g|}\ G_{b\partial}(\tz,\tx;\tx_1)\, G_{b\partial}(\tz,\tx;\tx_2)\, G_{b\partial}(\tz,\tx;\tx_3) + O(\lambda^2)\,,
\end{equation}
and similarly for other types of interaction. In \eqref{interactions-simplistic}, $G_{b\partial}(\tz,\tx;\tx_i)$ are the fuzzy bulk-to-boundary propagators evaluated between semi-classical states, for which we wrote integral representations in Section \ref{S:Two-point function and holography}. It would be interesting to obtain more explicit expressions for bulk-to-bulk and bulk-to-boundary propagators and use them to study correlators such as \eqref{interactions-simplistic}. It remains to be seen, though, if such expansions in modified Witten diagrams as above are the appropriate way to compute perturbative correlators on fuzzy AdS spaces, i.e. whether the Feynman-Witten rules follow from a consistent QFT based on an action.  We will study formal aspects of perturbative QFTs on fuzzy AdS spaces in future work.
\medskip

As a different approach to the above questions, one may proceed by a further analysis of boundary correlation functions and try to identify them either as correlators of some {\it deformed conformal field theory} or those of an ordinary CFT, but evaluated between states that are not conformally invariant. To this end, the first step will be to analyse the symmetries of the boundary two-point functions computed in this work. Naively, the fuzzy AdS$_2$ two-point function \eqref{fuzzy-2pt-AdS2-final} possesses none, not being even translationally invariant. We suspect, however, that symmetries are simply not manifest in position space. An interesting observation related to the second of the above possibilities is that the fuzzy AdS$_2$ boundary two-point function \eqref{fuzzy-2pt-AdS2-final} may be written as a three-point function
\begin{equation}\label{3pt-conclusions}
    G_{\partial\partial} = C_{\mathbb{\Delta}}\, \langle \phi(-\tau_1) \phi(\tau_2) \mathcal{D}\rangle\,,
\end{equation}
of an (ordinary) one-dimensional CFT. We have denoted $\tau_j = i\langle\hat t_j\rangle$ and $C_{\mathbb{\Delta}}$ is the normalisation constant given in \eqref{Joao-2pt-normalisation}. The operators entering the three-point function \eqref{3pt-conclusions} are two copies of a local primary operator $\phi$ of conformal dimension $\Delta_\phi = \mathbb{\Delta}/2-l$ and the non-local operator
\begin{equation}
    \mathcal{D} = \lambda_{\phi\phi\psi}^{-1} \frac{\Gamma(\mathbb{\Delta}-2l)^2}{\Gamma(-2l)^2\Gamma(2\mathbb{\Delta})}\int\limits_{-l\kbar}^\infty d\tau\, \tau^{-4l-1} \psi(\tau)\,,
\end{equation}
where $\psi$ is a primary field of dimension $2\Delta_\phi$ and $\lambda_{\phi\phi\psi}$ the OPE coefficient. E.g, if we take the CFT to be the generalised free theory of the fundamental scalar $\phi$, then $\psi = \phi^2$. The three-point function $\langle\phi(-\tau_1)\phi(\tau_2)\psi(\tau)\rangle$ is fixed by conformal symmetry and \eqref{3pt-conclusions} follows by integrating it over $\tau$.\footnote{On the other hand, in the limit $\kbar\to0$ with $l$ finite, not elaborated in the main text, the NC two-point function is closely related to a CFT two-point function in the presence of a {\it hyper-primary} operator, \cite{Khodaee:2017tbk}.} It would be interesting to understand the origin of the relation \eqref{3pt-conclusions} and determine if similar representations of fuzzy correlators exist for higher-point functions or for interacting theories.
\smallskip

The above perspective could also offer a way of studying dynamical gravity, as opposed to quantum field theory on a fixed background considered in this work (making the metric dynamical within the noncommutative frame formalism has met with difficulties -- see however \cite{Buric:2006di}). An interesting testing ground for this idea is provided by looking at fuzzy analogues of more interesting gravitational backgrounds, in particular of black hole solutions. A definition of the fuzzy BTZ black hole was already proposed in \cite{Buric:2022ton}. The boundary two-point function on this background, that should be computable along the lines of this work, is expected to give the finite-temperature two-point function of the above tentative 2d CFT. Putting all these facts into a consistent framework would be a long-term goal.
\medskip

An alternative approach to including dynamical gravity in our setup begins by noticing various similarities between fuzzy AdS spaces and {\it covariant quantum spaces} arising in the context of IKKT-like matrix models, \cite{Sperling:2018xrm,Sperling:2019xar,Steinacker:2024unq}. There seem to be some close relations between these two sets of noncommutative geometries which deserve further investigation. For instance, our semi-classical states can be used to define {\it quantum metric} and {\it quantum symplectic form} in the sense of \cite{Steinacker:2024unq}. If this is done, the corresponding symplectic volume is precisely the factor that appears in the resolution of the identity \eqref{resolution-of-the-identity}. On the other hand, the quantum metric closely resembles that of classical AdS space, but differs from it by certain relative signs between terms. An interesting further comparison that we will make in the future concerns the ‘deep quantum' regime of the two-point function. Namely, in the present work we computed expectation values of the two-point function inside pairs of semi-classical states, as opposed to more general matrix elements where the left and right pairs are not the same. In the terminology of \cite{Steinacker:2024unq}, this is the ‘semi-classical' regime. By contrast, in the ‘deep quantum' regime, the left and right pairs of states differ from each other, but for each pair the two semi-classical states are close to one another. For quantum spaces that originate from matrix models, it is expected on general grounds that the semi-classical and deep-quantum regimes are the only ones in which the two-point functions does not approximately vanish.
\smallskip

While these (existing or potential) connections are intriguing, their significance remains unclear. Nevertheless, it is tempting to try and realise our spaces as saddle points of some modified versions of IKKT-type matrix models.
\smallskip

A related set of questions, which is specific to three dimensions, regards the nature of internal degrees of freedom, such as their transformation properties. We saw that, when working in semi-classical states \eqref{semi-classical-states-AdS3}, the effect of the internal space was minimal, entering only through normalisation factors of modes. An approach to exploring the additional degrees of freedom would be through sets of suitably localised states which extend \eqref{semi-classical-states-AdS3}. One natural such family, provided by generalised coherent states in the sense of Perelomov, \cite{Perelomov:1986uhd}, is briefly discussed in Appendix \ref{A:Generalised coherent states in three dimensions}. It remains to be seen what other notions of coherent-like states are well-suited in the context of above models. See \cite{Manta:2025inq} for a related recent discussion.

\paragraph{The flat space limit and cosmology} We conclude by mentioning two extensions of the above constructions that may have interesting applications. In this work, we restricted ourselves to two and three dimensions, as well as a negative cosmological constant. The most relevant physical applications will require to drop either one or both of these restrictions. A particularly interesting case is that of AdS$_4$, for which most of the technology is already in place. Both the differential-geometric and representation-theoretic aspects of our construction are straightforwardly generalised to four dimensions. On the one hand, the coordinates can be defined as components of the Pauli-Lubanski vector of $\mathfrak{so}(3,2)$, akin to \cite{Buric:2017yes}, while momenta are given by translation and dilation generators. Furthermore, unitary representations of $\mathfrak{so}(3,2)$ are well-understood and many of them admit lowest-weight vectors. Therefore, the construction of semi-classical states would proceed in essentially the same way as in this work. Of course, it remains to be seen if the commutative limit with properties as above continues to exist in four dimensions. The AdS$_4$ is of particular interest in view of taking the flat space limit, $\ell_{\text{AdS}}\to\infty$. It is natural to speculate that this would lead to a quantum field theory on some noncommutative deformation of $\mathbb{R}^{1,3}$. In fact, the flat space limit and its properties present an interesting open question already for models studied in this work.
\smallskip

On the other hand, the case of positive cosmological constant, in particular the fuzzy dS$_4$, may be relevant for the early universe cosmology, where quantum gravitational effects play and important role. To this end, the main task that remains to be done is the definition of semi-classical states. It is still to be seen whether the principal or the discrete series representations of $SO(4,1)$ provide the more appropriate version of the fuzzy dS$_4$. In the case of the latter, one may speculate that semi-classical states can be defined as in this work, but this is to be further analysed. There are also some ideas for the appropriate notion of semi-classical states for principal series representations. Ultimately, the appropriateness of the set of states depends on having a good commutative limit. Assuming all the challenges are overcome, the appropriate analogue of Figure \ref{plot4} would be compared to inhomogeneities in the CMB. 

\section*{Acknowledgments}

The work of I.B. is funded by the Science Foundation Ireland under the grant agreement SFI-22/FFP-P/11444, and the work of B.B, M.B, D.D. and D.L. is supported by the Grant 451-03-136/2025-03/200162 of the Ministry of Science, Technological Development and Innovation of the Repiblic of Serbia. D.D. work was partially supported by the Science Fund of the Republic of Serbia, grant number TF C1389-YF, “Towards a Holographic Description of Noncommutative Spacetime: Insights from Chern-Simons Gravity, Black Holes and Quantum Information Theory" -- HINT.

\appendix

\section{Classical propagator as a sum over modes}
\label{A:Classical propagator as a sum over modes}

In this appendix, we review how the AdS$_{d+1}$ propagator \eqref{classical-bulk-propagator} arises as an integral over field modes, focusing on $d=2,1$. Among the several two-point functions that may be studied, we consider the vacuum expectation value of the ordinary product of field operators
\begin{equation}\label{choice-of-two-pt-function}
    G_{bb}^{\text{comm}} = \langle0|\Phi(\tx_1,\tz_1)\Phi(\tx_2,\tz_2)|0\rangle\ .
\end{equation}
In terms of field modes, the latter expands as
\begin{equation}
    G_{bb}^{\text{comm}} = \int \diff \omega\, \diff ^{d-1} k \, \phi_{\omega,k}(\tx_1,\tz_1) \,\phi^\ast_{\omega,k}(\tx_2,\tz_2)\ .
\end{equation}
The modes on AdS$_3$ read
\begin{equation}\label{classical.modes}
    \phi_{\omega,k}(\tx,\tz) = \frac{\tz}{\sqrt{4\pi}} \,e^{-i(\omega\t - k\tx) }J_{\nu}\left(\sqrt{\omega^2-k^2}\, \tz\right)\,,
\end{equation}
and therefore, we may write the two-point function as
\begin{equation}
    G_{bb}^{\text{comm}} = \frac{\tz_1\tz_2}{4\pi}\int\limits_0^\infty \diff \omega\, \int\limits_{-\omega}^{\omega}\diff k \,e^{-i\omega(\t_1-\t_2)+ik(\tx_1-\tx_2)}J_{\nu}\left(\sqrt{\omega^2-k^2}\, \tz_1\right)J_{\nu}\left(\sqrt{\omega^2-k^2}\, \tz_2\right) \ .
\end{equation}
To make progress, we make the change of variables
\begin{equation}
    \omega = \Omega\cosh{\theta}\,, \qquad k = \Omega\, \sinh{\theta}\,,
\end{equation}
which gives
\begin{equation}
     G_{bb}^{\text{comm}} = \frac{\tz_1\tz_2}{4\pi}\int\limits_0^\infty \diff \Omega\, \Omega \, J_\nu(\Omega \tz_1) J_\nu(\Omega \tz_2) \int\limits_{-\infty}^\infty \diff \theta \, e^{-i\Omega\,(\t_1-\t_2)\,\cosh{\theta}+i\Omega\,(\tx_1-\tx_2)\,\sinh{\theta}}\ .
\end{equation}
Let us denote $\t_{12} = \t_1-\t_2$ and $\tx_{12} = \tx_1-\tx_2$. To regularise and evaluate the integral over $\theta$, we use the formula 10.9.15 from \cite{NIST:DLMF} (with $\nu=0$)
\begin{equation}\label{nist1}
    \int\limits_{-\infty}^\infty \diff\theta\ e^{iz\cosh \theta + i\zeta \sinh \theta} = \pi i H^{(1)}_0\left(\sqrt{z^2-\zeta^2}\right)\,, \qquad \text{Im}(z\pm\zeta)>0\ .
\end{equation}
The $\epsilon$-prescription we use is to first set $z = -\Omega\t_{12}+ i\epsilon$ and $\zeta=\Omega\tx_{12}$, then apply \eqref{nist1} and finally put $\epsilon=0$. Further using the relation between Hankel functions and modified Bessel functions, we get
\begin{equation}
    G_{bb}^{\text{comm}} = \frac{\tz_1\tz_2}{2\pi}
    \int\limits_0^{\infty}\diff \Omega\, \Omega \, J_{\nu}(\Omega \tz_1)\, J_{\nu}(\Omega \tz_2)\,K_0\left(\Omega\sqrt{\tx_{12}^2-\t_{12}^2}\right)\ .
\end{equation}
The last integral may be evaluated using the formula 6.522 3 of \cite{Gradshteyn}, yielding 
\begin{equation}
    G_{bb}^{\text{comm}} =\frac{1}{2\pi} \frac{2^{\nu}}{\sqrt{\zeta(\zeta +4)}} \left(\zeta+2+\sqrt{\zeta(\zeta+4)}\right)^{-\nu}.
\end{equation}
This coincides with the expression \eqref{classical-bulk-propagator} upon putting in the latter $d=2$ and using the special hypergeometric identity 15.4.8 of \cite{NIST:DLMF}.
\smallskip 

Similarly, on AdS$_2$, the propagator is the integral over the modes
\begin{equation}
    G_{bb}^{\text{comm}} = \int\limits_0^{\infty}\diff \omega\, \phi_{\omega}(\t_1,\tx_1) \,\phi^\ast_{\omega}(\t_2,\tx_2) = \frac12\sqrt{\tz_1\tz_2}  \int\limits_0^{\infty}\diff \omega\, e^{-i\t_{12}} J_\nu(\omega\tz_1) J_\nu(\omega\tz_2)\ .
\end{equation}
Again, the integral requires regularisation. We will use the formula 6.612 3 of \cite{Gradshteyn} with
\begin{equation}
    \alpha = i\t_{12} + \epsilon\,, \qquad \beta = \tz_1\,, \qquad \gamma = \tz_2\,,
\end{equation}
and subsequently set $\epsilon$ to zero. This process gives the two-point function
\begin{equation}\label{AdS2-classical-propagator}
    G_{bb}^{\text{comm}} = \frac{1}{2\pi}Q_{\DDelta-1}\left(1+\frac{\zeta}{2}\right) = \frac{1}{2\pi} \frac{\Gamma^2(\DDelta)}{\Gamma(2\DDelta)}\frac{2^{2\DDelta-1}}{\zeta^\DDelta}\ _2F_1\left(\DDelta,\DDelta;2\DDelta;\frac{-4}{\zeta}\right) \ .
\end{equation}
In the last step, we have used the identity 14.3.19 of \cite{NIST:DLMF} that relates Legendre and hypergeometric functions,
\begin{equation}
    Q_{\DDelta-1}(x)=\frac{\Gamma^2(\DDelta)}{\Gamma(2\DDelta)}\,\frac{2^{\DDelta-1}}{(x-1)^\DDelta}\ _2F_1\left(\DDelta,\DDelta;2\DDelta;\frac{2}{1-x}\right)\ .
\end{equation}
In \eqref{AdS2-classical-propagator} we recognise the two-point function \eqref{classical-bulk-propagator}.

\section{Representations of $SL(2,\mathbb{R})$}
\label{A:Representations}

In this appendix, we collect some properties of the group $SL(2,\mathbb{R})$ and its representations that are used in the main text. The Lie algebra $\mathfrak{sl}(2,\mathbb{R})$ is spanned by generators $\{H,E_\pm\}$ that obey the brackets
\begin{equation}
    [H,E_+] = E_+\,, \quad [H,E_-] = -E_-\,, \quad [E_+,E_-] = 2H\ .
\end{equation}
We often make use of the linear combinations of these generators,
\begin{equation}
    \tilde{H}=\frac{i}{2}(E_+-E_-)\,, \qquad \tilde{E}_+=\frac{1}{2}(E_+ +E_- +2iH)\,, \qquad \tilde{E}_-=\frac{1}{2}(E_++E_--2iH)\,,
\end{equation}
which satisfy the same brackets as $\{H,E_\pm\}$. These linear combinations are convenient because $\tilde H$ generates the maximal compact subgroup of $SL(2,\mathbb{R})$, and thus has a discrete spectrum in any unitary irreducible representation.
\smallskip

The group $SL(2,\mathbb{R})$ has two families of discrete series representations, denoted $T_l^-$ and $T_l^+$, \cite{Vilenkin}. Here, $l$ is an integer or a half-integer and for the first family $l\leq-1$, while $l\geq 1$ for the second family. The carrier space of $T_l^{-}$ will be denoted by $\mathcal{H}_l$. The two types of representations are related by
\begin{equation}
    \left(T_l^-\right)^\ast = T_{-l}^+\,,
\end{equation}
where $\pi^\ast$ stands for the dual (contragredient) representation to $\pi$. Another way to relate the positive and negative discrete series is by means of the {\it Bargmann automorphism}, $B$, i.e. $\left(T_l^-\right)^B=T_{-l}^+$. The Bargmann automorphism is an outer automorphism of $SL(2,\mathbb{R})$ that acts on the generators by
\begin{equation}\label{Bargmann-automorphism}
     B(H)=H\,, \qquad B(E_+)=-E_+\,, \qquad B(E_-)=-E_-\ .
\end{equation}
The negative discrete series representations are of lowest-weight type. The lowest-weight vector $\Psi_0$ satisfies
\begin{equation}
    \tilde{E}_- \Psi_0=0\,, \qquad \tilde{H}\Psi_0=-l\Psi_0\,,
\end{equation}
and the representation space is spanned by vectors $\Psi_n = \tilde E_+^n \Psi_0$. 

We shall use the realisation of $T_l^-$ on the space of functions $f(\xi)$, with $\xi>0$, square-integrable with respect to the inner product,
\begin{equation}\label{inner-product-discrete-series}
    (f_1,f_2) = 2^{2l+1}\pi \int\limits_0^\infty \diff \xi \,\xi^{2l+1} f_1(\xi)^\ast f_2(\xi)\ .
\end{equation}
The action of generators on these functions reads
\begin{equation}
    H=\xi \partial_\xi + l+1\,, \qquad E_+=-i\xi\,, \qquad  E_- = -i\left( \xi\partial_\xi^2 + 2 (l+1) \partial_\xi\right)\ .
\end{equation}
We use the notation $f(\xi) = \langle \xi|f\rangle$; due to the nontrivial measure in the scalar product in \eqref{inner-product-discrete-series}, it is then consistent to write
\begin{equation}
    \langle\xi|\xi'\rangle = (\pi\,2^{2l+1}\xi^{2l+1})^{-1}\,\delta(\xi-\xi')\,, 
\end{equation}
where $\delta(\xi-\xi')$ is the standard $\delta$-function. The `ket' $|\xi\rangle$ is proportional to the eigenvector of  $\hz\,$,
\begin{equation}
  \langle\xi|z\rangle =(\kbar\pi\,2^{2l+1}\xi^{2l+1})^{-1/2}\,\delta\left(\xi-\frac{z}{\kbar}\right)\,,
\end{equation}
while the eigenfunctions of the time $\hat t$ are
\begin{equation}
  \langle\xi|t\rangle =(\kbar^{1/2}\pi\,2^{l+1})^{-1}\,\xi^{-l-1+i\, \frac t\kbar} \ .
\end{equation}
The lowest-weight vector is given by the function
\begin{equation}
    \Psi_0 (\xi) = \frac{\xi^{-2l-1} e^{-\xi}}{2^{2l+1/2}\sqrt{\pi\Gamma(-2l)}} \equiv N \xi^{-2l-1} e^{-\xi}\ .
\end{equation}

\subsection{Representations of $SO(2,2)$}

The Lie algebra $\mathfrak{so}(2,2)$ is isomorphic to the direct sum of two copies of $\mathfrak{sl}(2,\mathbb{R})$,
\begin{equation}
    \mathfrak{so}(2,2) \cong \mathfrak{sl}(2,\mathbb{R}) \oplus \mathfrak{sl}(2,\mathbb{R})\ .
\end{equation}
Therefore, its representations take the form of tensor products $\pi_1\otimes\pi_2$ of representations $\pi_i$ of $\mathfrak{sl}(2,\mathbb{R})$. It is common to distinguish between the two factors by putting the bars on all quantities connected with the second one. Following this convention, we shall denote the tensor product of two copies of the discrete series representation $T_l^-$ by $T_l^- \otimes \bar T_l^-$. This is naturally realised on the space of functions in two variables $f(\xi,\bar\xi)$ and has the lowest-weight vector
\begin{equation}
    \Psi_0 \otimes \bar \Psi_0 = N^2 (\xi\bar\xi)^{-2l-1} e^{-\xi-\bar\xi}\ .
\end{equation}

\section{Field modes on the fuzzy AdS$_3$}
\label{A:Field modes on the fuzzy AdS$_3$}

In this appendix, we derive the expectation values of field modes on the fuzzy AdS$_3$ in semi-classical states, \eqref{modes-AdS3-1}. In coordinates $(\rho,\varphi,\zeta_L,\zeta_R)$ from Section \ref{SS:Solutions to the Klein-Gordon equation}, and using the expression \eqref{coherent-states-ads3-rep}, the integral \eqref{modes-AdS3-1} is re-written as
\begin{align}
    &\langle\phi_\gamma\rangle = 2 N^4 \left(2^{2l+1}\pi\right)^4 c_\gamma \iiiint \diff\rho \,\diff\varphi \,\diff\zeta_+\, \diff\zeta_-\ \lambda^{4l} \rho^{-4l} \sin(2\varphi)\, e^{-\frac{2\rho}{\lambda}f_\lambda + \frac{2i\rho}{\lambda}(b f_b + c f_c)}\\
    & \hskip6cm \times J_\nu(2\gamma\rho) \delta(\varphi-\varphi_0)\,\delta(\zeta_+-\zeta_{+0})\,\delta(\zeta_--\zeta_{-0})\,, \nonumber
\end{align}
with
\begin{align}
   &  f_\lambda = \cos\varphi\cosh\zeta_L + \sin\varphi\cosh\zeta_R\,,\label{fs-1}  \\
   &  f_b = \cos\varphi\cosh\zeta_L - \sin\varphi\cosh\zeta_R\,, \label{fs-2}\\
   &  f_c = \cos\varphi\sinh\zeta_L - \sin\varphi\sinh\zeta_R\ . \label{fs-3} 
\end{align}
The integrals over delta functions are readily performed, leaving us with
\begin{equation}
    \langle\phi_\gamma\rangle = 2 N^4 \left(2^{2l+1}\pi\right)^4 c_\gamma \lambda^{4l} \sin(2\varphi_0) \int \diff \rho\ \rho^{-4l} \, e^{-\frac{2\rho}{\lambda}f_{\lambda 0} + \frac{2i\rho}{\lambda}(b f_{b0 }+ c f_{c0})} J_\nu(2\gamma\rho)\ .
\end{equation}
For the integral over $\rho$, we use the identity \eqref{GR-integral}, with parameters
\begin{equation}\label{alpha-const}
    \mu=1-4l\,, \qquad \beta = 2\gamma\,, \qquad \alpha = \frac{ \ 2 f_{\lambda 0} - 2i (b f_{b0 }+ c f_{c0})}{\lambda}\ .
\end{equation}
We see that for sufficiently small $\lambda$, the conditions \eqref{conditions} are satisfied. We obtain
\begin{align}\label{modes-ads3-appendix-C}
    \langle\phi_\gamma\rangle = 2^{8l+5} \pi^4 N^4 c_\gamma \lambda^{4l} \sin2\varphi_0\, &\frac{\Gamma(\nu+1-4l)}{\Gamma(\nu+1)} \, \frac{ \gamma^\nu}{\alpha^{\nu+1-4l}}\\
    & \times \, _2F_1\left(\frac{\nu+1-4l}{2},\frac{\nu+2-4l}{2};\nu+1;-\frac{4\gamma^2}{\alpha^2}   \right)\ . \nonumber
\end{align}
We wish to re-express the result in terms of expectation values of coordinates and take the $l\to-\infty$ limit. In this limit, the expectation values \eqref{coord-expectation-ads3-1}-\eqref{coord-expectation-ads3-2} reduce to
\begin{equation}
      \langle \hat z\rangle \sim -2l\kbar \lambda\,, \qquad \langle \hat t\rangle \sim -2l\kbar c\,, \qquad \langle\hat x\rangle \sim -2l\kbar b\ .
\end{equation}
On the other hand, the variable $\alpha$ reads
\begin{equation}
    \alpha = \frac{-4l\kbar f_{\lambda 0}}{\langle\hz\rangle}\left( 1 + \frac{i}{2l\kbar} \left(\frac{f_{b0}}{f_{\lambda0}} \langle\hx\rangle + \frac{f_{c0}}{f_{\lambda0}} \langle\hat t\rangle \right) \right) \equiv \frac{-4l\kbar f_{\lambda0}}{\langle\hz\rangle}\left( 1 - \frac{i\omega\langle\hat t\rangle}{4l} + \frac{ik\langle\hx\rangle}{4l} \right)\,,
\end{equation}
where, anticipating the commutative limit, we have defined
\begin{align}\label{omega-k-definition}
   \omega =- \frac{2}{\kbar} \frac{f_{c0}}{f_{\lambda0}}\ ,\qquad 
  k = \frac{2}{\kbar} \frac{f_{b0}}{f_{\lambda0}} 
 \ .
\end{align}
To take the commutative limit, $l\to-\infty$, we use the formula \eqref{2F1-Bessel-limit} with
\begin{equation}
    a = -2l + \frac{\nu+1}{2}\,, \qquad b = -2l + \frac{\nu+2}{2}, \qquad \tilde z^2 = 16ab \,\frac{\gamma^2}{\alpha^2}\ .
\end{equation}
In the commutative limit, $\tilde z$ remains finite and tends to
\begin{equation}
    \lim_{l\to-\infty}\tilde z = \frac{2\gamma \langle\hz\rangle}{\kbar f_{\lambda0}} = \sqrt{\omega^2 - k^2} \, \langle\hz\rangle\ .
\end{equation}
Putting everything together,
\begin{align}
    \langle\phi_\gamma\rangle = \frac{2^{4l+3} \pi^2}{\Gamma(-2l)^2\ } \, c_\gamma \,\sin2\varphi_0 &\, \frac{\Gamma(\nu+1-4l)}{(-4l)^{\nu+1}} \left( 1 - \frac{i\omega\langle\hat t\rangle}{4l} +\frac{ik\langle\hx\rangle}{4l} \right)^{4l-\nu-1}\\
    &\hskip2cm \times  \left(\frac{ 1}{f_{\lambda 0}}\right)^{1-4l} \kbar^{-1}\langle\hz\rangle J_\nu\left(\sqrt{\omega^2 - k^2}\,\langle\hz\rangle \right)\ . \nonumber
\end{align}
Finally, upon recognising the limit that is the definition of the exponential function, we obtain the commutative field modes written in \eqref{modes-ads3-classical-limit},
\begin{equation}
    \langle\phi_\gamma\rangle = 2^{\frac72} \pi^2 (-l)^{\frac12} \kbar^{-1} c_\gamma \left(f_{\lambda 0}\right)^{4l-1}\sin2\varphi_0\ \frac{1}{\sqrt{4\pi}}\, e^{-i\omega\langle\hat t\rangle + ik\langle\hx\rangle} \langle\hz\rangle \, J_\nu\left(\sqrt{\omega^2 - k^2}\,\langle\hz\rangle \right)\ .
\end{equation}

\paragraph{Remark} Before the commutative limit, the variable $\alpha$ reads
\begin{equation}
    \alpha = \frac{-4l\kbar f_{\lambda0}}{\langle\hz\rangle}\left( \frac{\Gamma(1/2-2l)^2}{-2l\Gamma(-2l)^2} - \frac{i\omega\langle\hat t\rangle}{4l} + \frac{ik\langle\hx\rangle}{4l} \right) 
\end{equation}
Using this equation and the relation
\begin{equation}
    \gamma^2 = f_{c0}^2 - f_{b0}^2 = \frac{\hbar^2}{4} f_{\lambda0}^2 (\omega^2 - k^2)\,,
\end{equation}
we see that the argument of the hypergeometric appearing in \eqref{modes-ads3-appendix-C} only depends on the quantum numbers $\omega$ and $k$. Therefore, the dependence of the expectation values \eqref{modes-ads3-appendix-C} of the mode on the third quantum number, even before taking the commutative limit, is only through overall normalisation.

\section{Appell functions}

Appell functions are certain generalisations of the hypergeometric function to functions of two variables. There are four of them, $F_1,\dots, F_4$, and they are well integrated in \textit{Mathematica}. We briefly summarise the definition and some properties of $F_1$, the function that appears in the main text. For a comprehensive account, see \cite{Appell}.
\smallskip

The Appell function $F_1$ may be defined through its double series expansion
\begin{equation}
    F_1(\alpha,\beta,\beta';\gamma;x,y) =\sum_{m,n=0}^\infty \frac{(\alpha)_{m+n} (\beta)_m (\beta')_n}{(\gamma)_{m+n} m! \,n!}\ x^m y^n \,,\qquad |x|,|y|<1  \ ,
\end{equation}
where $(\alpha)_n$ is the Pochhammer symbol. It is clearly symmetric under the simultaneous exchange $\beta\leftrightarrow \beta'$, $x\leftrightarrow y$. The function may also be characterised through a pair of differential equations that it satisfies. One of these equations reads
\begin{equation}
    x(1-x)\,\frac{\p^2 F_1}{\p x^2} + y(1-x)\, \frac{\p^2F_1}{\p x\,\p y}+ \big( \gamma-(\alpha+\beta+1) x\big) \,\frac{\p F_1}{\p x} -\beta y\, \frac{\p F_1}{\p y} - \alpha\beta\, F_1 =0 \,,
\end{equation}
and the second is obtained by $\beta\leftrightarrow \beta'$, $x\leftrightarrow y$. As the hypergeometric function, $F_1$ obeys a number of relations; for particular values of variables or parameters it reduces to the hypergeometric function. The two properties used in this work are integral representation
\begin{equation}
    F_1(\alpha,\beta,\beta';\gamma;x,y) = \frac{\Gamma(\gamma)}{\Gamma(\alpha)\Gamma(\gamma-\alpha)} \int\limits_0^1 \diff t \ t^{\alpha-1} (1-t)^{\gamma-\alpha-1} (1 - x t)^{-\beta} (1 - y t)^{-\beta'}\,, 
\end{equation}
valid for $\text{Re}(\gamma) > \text{Re}(\alpha)>0$, and the reduction formula
\begin{equation}
    \lim_{\varepsilon\to 0} F_1\left(\alpha,\frac 1\varepsilon,\frac 1\varepsilon;\gamma;\,\varepsilon x,\,\varepsilon y\right) =\ _1F_1(\alpha,\gamma,x+y) \ .
\end{equation}
Here, $_1F_1(\alpha,\gamma,z) $ is Kummer's confluent hypergeometric function.

\section{Generalised coherent states in three dimensions}
\label{A:Generalised coherent states in three dimensions}

In the main text, we focused on semi-classical states \eqref{semi-classical-states-AdS2}, \eqref{semi-classical-states-AdS3}, in which we computed the expectation values of field modes and the propagator. These states are defined by means of momenta, and hence by the noncommutative frame. In two dimensions, they coincide with generalised coherent states in the sense of Perelomov, \cite{Perelomov:1986uhd}. However, in three dimensions, where there are additional internal degrees of freedom, our semi-classical states \eqref{semi-classical-states-AdS3} are only a subset of generalised coherent states of \cite{Perelomov:1986uhd}, the latter providing a one-parameter extension of the former. It is therefore tempting to try and use the full set of generalised coherent states in order to probe properties of internal degrees of freedom. The present appendix gives some initial steps in this direction.
\smallskip

Following the general theory of \cite{Perelomov:1986uhd}, generalised coherent states associated to the representation space $\mathcal{H} = \mathcal{H}_l \otimes\mathcal{\bar H}_l$ (the algebra of fuzzy functions \eqref{fuzzy-algebra-ads3} being $\mathcal{A} = \text{End}(\mathcal{H})$) read
\begin{equation}\label{quasi-coherent-states}
    |\mu,\lambda,b,c\rangle = e^{\mu(\bar H - H)} \lambda^{-(H+\bar H)} e^{-b (E_+ - \bar E_+)}  e^{-c (E_+ + \bar E_+)} |\Psi_0\otimes\bar\Psi_0\rangle\ .
\end{equation}
The order of the exponentials and the precise linear combinations of generators $H,\bar H,E_+,\bar E_+$ entering them are a matter of convention. Choices other than \eqref{quasi-coherent-states} correspond to changes of coordinates $(\mu,\lambda,b,c)$. Recalling our definition of momenta $\hat p_z,\hat p_t, \hat p_x$, equations \eqref{coord-momenta-ads3-1}-\eqref{coord-momenta-ads3-3}, we see that for $\mu=0$, \eqref{quasi-coherent-states} are precisely the the semi-classical states \eqref{semi-classical-states-AdS3}. In the realisation of the discrete series that we use, generalised coherent states read
\begin{equation}\label{extended-coherent-states-explicit}
    \langle\xi,\bar\xi | \mu,\lambda,b,c\rangle = N^2 \lambda^{2l} (\xi\bar\xi)^{-2l-1} e^{\frac{i}{\lambda}\left(e^{-\mu} \xi (b+c+i) + e^\mu \bar\xi (b-c+i)\right)}\ .
\end{equation}
The states \eqref{quasi-coherent-states} give a resolution of the identity. This follows from abstract arguments, but can also be verified explicitly using \eqref{extended-coherent-states-explicit}. Indeed, putting
\begin{equation}
    P_{\text{AdS}_3} = \int\limits_{-\infty}^\infty \mathrm{d}\mu \int\limits_{-\infty}^\infty \mathrm{d}b \int\limits_{-\infty}^\infty \mathrm{d}c \int\limits_0^\infty \frac{\mathrm{d}\lambda}{\lambda} \ |\mu,\lambda,b,c\rangle \langle\mu,\lambda,b,c|\,,
\end{equation}
and working with the form \eqref{extended-coherent-states-explicit} of the states, one obtains
\begin{equation}\label{completeness-AdS3-1}
    \langle \xi,\bar\xi | P_{\text{AdS}_3} | \xi',\bar\xi' \rangle = \frac{2^{-4l}}{(2l+1)^2} \left(\xi \bar\xi\right)^{-2l-1} \delta(\xi - \xi') \delta(\bar\xi - \bar\xi')\ .
\end{equation}
The derivation of \eqref{completeness-AdS3-1} is similar to that of \eqref{resolution-of-the-identity} and makes use of the integrals
\begin{align}
     \int\limits_{-\infty}^\infty \mathrm{d}\mu\ e^{- \frac{2\xi}{\lambda} e^{-\mu} - \frac{2\bar\xi}{\lambda} e^\mu} &= 2 K_0\left(\frac{4\sqrt{\xi\bar\xi}}{\lambda}\right)\,,\\
     \int\limits_0^\infty \mathrm{d}\lambda\ \lambda^{4l+1} K_0 \left(\frac{4\sqrt{\xi\bar\xi}}{\lambda}\right) & = 16^l \Gamma(-2l-1)^2 (\xi\bar\xi)^{2l+1}\,,
\end{align}
where $K_0$ is the modified Bessel function of the second kind. Recalling the inner product \eqref{inner-product-discrete-series} in the discrete series, the statement \eqref{completeness-AdS3-1} is precisely the resolution of the identity. On the other hand, the semi-classical states $\,\{|\mu=0, \lambda,b,c\rangle \}$ on their own do not resolve the identity. Setting
\begin{equation}
    P_{\text{AdS}_3}^0 = \int\limits_{-\infty}^\infty \mathrm{d}b \int\limits_{-\infty}^\infty \mathrm{d}c \int\limits_0^\infty \frac{\mathrm{d}\lambda}{\lambda} \ |\lambda,b,c\rangle \langle\lambda,b,c|\,,
\end{equation}
one finds
\begin{equation}
    \langle\xi,\bar\xi| P_{\text{AdS}_3}^0 | \xi',\bar\xi' \rangle = \frac{2^{-4l+1}\Gamma(-4l-2)}{\Gamma(-2l)^2}\left(\frac{\xi+\bar\xi}{\xi\bar\xi}\right)^{4l+2} \delta(\xi - \xi') \delta(\bar\xi - \bar\xi')\ .
\end{equation}
\medskip

With the generalised coherent states $|\mu,\lambda,b,c\rangle$, one may start probing the geometry of the internal space. Notice that, on the commutative AdS$_3$, the operator $\bar H - H$ generates boosts in the $\t-\tx$ plane, while preserving the coordinate $\tz$. The analogous statements also hold in the fuzzy case: denoting expectation values of operators $\mathcal{O}$ in states \eqref{quasi-coherent-states} by $\langle\mathcal{O}\rangle$, we have
\begin{equation}
    \langle \hat z \rangle = h\lambda\,, \qquad  \langle \hat t \rangle = h \left(c \cosh\mu - b \sinh\mu\right) \,, \qquad \langle \hat x \rangle = h\left( b \cosh\mu - c \sinh\mu\right)\ ,
\end{equation}
with $h$ given by
\begin{equation}
  h  = \frac{\Gamma\left(\frac12-2l\right)^2 }{\Gamma(-2l)^2}\,\kbar \ .
\end{equation}
These relations follow from the Baker-Campbell-Hausdorff formula. The expectation values of field modes in generalised coherent states read
\begin{align}\label{modes-expectation-quasi-coherent}
    \langle\phi_\gamma\rangle = 2^{8l+5} \pi^4 N^4 c_\gamma \lambda^{4l} \sin2\varphi_0\, &\frac{\Gamma(\nu+1-4l)}{\Gamma(\nu+1)} \, \frac{ \gamma^\nu}{\alpha(\mu)^{\nu+1-4l}}\\
    & \times \, _2F_1\left(\frac{\nu+1-4l}{2},\frac{\nu+2-4l}{2};\nu+1;-\frac{4\gamma^2}{\alpha(\mu)^2}   \right)\,, \nonumber
\end{align}
generalising \eqref{modes-ads3-appendix-C}. Here $\alpha(\mu)$ is defined in the same way as above
\begin{equation}\label{alpha-const-aE}
    \alpha(\mu) = \frac{ \ 2 f_{\lambda 0}(\mu) - 2i (b f_{b0}(\mu) + c f_{c0}(\mu))}{\lambda}\,,
\end{equation}
however with $f_\lambda,f_b,f_c$ having shifted arguments, $\zeta_L\to\zeta_L-\mu$, $\zeta_R \to \zeta_R - \mu$, (that is, $\zeta_-\to\zeta_-$,  $\zeta_+\to\zeta_+-\mu$) :
\begin{align}
   &  f_\lambda(\mu) = \cos\varphi\cosh(\zeta_L-\mu) + \sin\varphi\cosh(\zeta_R-\mu)\,, \\
   &  f_b(\mu) = \cos\varphi\cosh(\zeta_L-\mu) - \sin\varphi\cosh(\zeta_R-\mu)\,, \\
   &  f_c(\mu) = \cos\varphi\sinh(\zeta_L-\mu) - \sin\varphi\sinh(\zeta_R-\mu)\ .
\end{align}
If we define, in analogy with \eqref{coord-momenta-ads3-1} - \eqref{coord-momenta-ads3-3} and \eqref{coordinates.limit}, $\mu=\kbar\sigma/2h$, and take the limit $l\to-\infty$ in \eqref{modes-expectation-quasi-coherent}, we get
\begin{equation}\label{modes-in-quasi-coherent-l-infinity}
    \lim_{l\to-\infty} \langle\phi_\gamma\rangle = \frac{1}{\pi\sqrt{2}} \,e^{\mathfrak{s}\sigma}  e^{-i\omega\langle\hat t\rangle + i k \langle\hat x\rangle } \langle\hat z\rangle J_\nu \left(\sqrt{\omega^2 - k^2}\langle\hat z\rangle\right)\,,
\end{equation}
where
\begin{align}\label{s-definition}
    \mathfrak{s} & = \frac{\cos\varphi_0\sinh\zeta_L + \sin\varphi_0\sinh\zeta_R}{\cos\varphi_0\cosh\zeta_L + \sin\varphi_0\cosh\zeta_R}\\[4pt]
    & = \frac{-2\kbar\omega + \sqrt{2} \sqrt{4+\kbar^2\left(k^2+\omega^2\right) - 8\kbar k \cos2\varphi_0 + \left(4+\kbar^2 \left(k^2-\omega^2\right)\right)\cos 4\varphi_0}}{4\cos2\varphi_0}\ \ . \nonumber
\end{align}
From the first line \eqref{s-definition}, we see that $\mathfrak{s}\in(0,1)$. In particular, the zero of the denominator at $\varphi_0=\pi/4$ is cancelled by the numerator which also vanishes at this point.
\smallskip

Let us end with some preliminary remarks on the interpretation of \eqref{modes-in-quasi-coherent-l-infinity}. We have worked in a basis of modes corresponding to $\delta$-functions in the $\varphi$-variable. We can switch to a discrete basis by Fourier expanding \eqref{modes-in-quasi-coherent-l-infinity} using \eqref{s-definition}. It is slightly simpler to expand in powers $\cos^n(2\varphi)$ rather than in $\cos(2n\varphi)$. In this basis, the first two modes read
\begin{align}\label{order-0-boost}
    \lim_{l\to-\infty} \langle\phi_\gamma^{(n=0)}\rangle & = \frac{e^{-\frac{k}{\omega}\sigma}}{\pi\sqrt{2}} \, e^{-i\omega\langle\hat t\rangle + i k \langle\hat x\rangle } \langle\hat z\rangle J_\nu \left(\sqrt{\omega^2 - k^2}\langle\hat z\rangle\right)\,,\\
    \lim_{l\to-\infty} \langle\phi_\gamma^{(n=1)}\rangle & = \frac{e^{-\frac{k}{\omega}\sigma}}{\pi\sqrt{2}} \, \frac{(\omega^2-k^2)(4-\kbar^2\omega^2)}{4\kbar\omega^3}\sigma e^{-i\omega\langle\hat t\rangle + i k \langle\hat x\rangle } \langle\hat z\rangle J_\nu \left(\sqrt{\omega^2 - k^2}\langle\hat z\rangle\right)\,,\label{order-1-boost}
\end{align}
and similarly for the higher ones. Equations \eqref{order-0-boost}-\eqref{order-1-boost} indicate the behaviour of different field modes under boosts in the $t-x$ plane (recall that the boost generator $\bar H - H$ commutes with the fuzzy Laplacian, so it is a symmetry of the noncommutative theory). If we take the classical limit $\kbar\to0$ in \eqref{modes-in-quasi-coherent-l-infinity}, keeping $\sigma$ constant, the $\varphi$-dependence trivialises and we obtain
\begin{equation}\label{full-commutative-limit-generalised-coherent}
    \lim_{\kbar\to0}\lim_{l\to-\infty} \langle\phi_\gamma\rangle = \frac{e^\sigma}{\pi\sqrt{2}}\, e^{-i\omega\langle\hat t\rangle + i k \langle\hat x\rangle } \langle\hat z\rangle J_\nu \left(\sqrt{\omega^2 - k^2}\langle\hat z\rangle\right)\ .
\end{equation}
We leave the interpretation of \eqref{order-0-boost}-\eqref{full-commutative-limit-generalised-coherent}, as well as further studies of different families of localised states in three dimensions, for future work.

\bibliographystyle{JHEP}
\bibliography{bibliography}

\end{document}